%% file: aaaaa.tex
\documentclass[longauth]{aa} 

\usepackage{graphicx}
\usepackage{txfonts}
\usepackage[dvipsnames]{xcolor}
\usepackage{hyperref}
\hypersetup{
    colorlinks = true,
    linkbordercolor = white,
    citecolor = blue,
    linkcolor=blue,
    urlcolor = blue
}
\usepackage{amsmath}	
\usepackage{amssymb}	

\usepackage{longtable}  
\usepackage{xtab,booktabs}   %
\usepackage{threeparttable}  

    \usepackage{CJK}

\definecolor{mviola}{rgb}{0.54, 0.16, 0.8}
\definecolor{molive}{rgb}{0.33, 0.42, 0.18}
\definecolor{mgrey}{rgb}{0.1, 0.1, 0.3}

\newcommand{\swift}{\textit{Swift}}
\newcommand{\kw}{Konus-\textit{Wind}}

\begin{document} 
\begin{CJK*}{UTF8}{gbsn}
\title{A blast from the infant Universe: the very high-$z$ GRB 210905A
\thanks{Based on observations collected at the Very Large Telescope of the European Southern Observatory,  Paranal, Chile (ESO programme 106.21T6; PI: N. Tanvir), the Hubble Space Telescope (programme 16918; PI: N. Tanvir),  REM (AOT43; programme  43008; PI: A. Melandri), and GROND (0106.A-9099(A); PI: A. Rau).}
}
\titlerunning{The very high-$z$ GRB 210905A}

\input{authaa.tex}

\date{Received xx, xx; accepted xx, xx}

\abstract{
We present a detailed follow-up of the very energetic GRB 210905A at a high redshift of $z=6.312$ and its luminous X-ray and optical afterglow.
Following the detection by \swift\ and \kw, we obtained a photometric and spectroscopic follow-up in the optical and near-infrared (NIR), covering both the prompt and afterglow emission from a few minutes up to 20 Ms after burst.
With an isotropic gamma-ray energy release of $E_\mathrm{iso} = 1.27_{-0.19}^{+0.20} \times 10^{54}$~erg, GRB 210905A lies in the top $\sim7\%$ of gamma-ray bursts (GRBs) in the \kw{} catalogue in terms of energy released. Its afterglow is among the most luminous ever observed, and, in particular, it is one of the most luminous in the optical at $t\gtrsim0.5$ d in the rest frame. 
The afterglow starts with a shallow evolution that can be
explained by energy injection, and it is followed by a steeper decay, while the spectral energy distribution is in agreement with slow cooling in a constant-density environment within the standard fireball theory. A jet break at $\sim46.2\pm16.3$ d ($6.3\pm2.2$ d rest-frame) has been observed in the X-ray light curve; however, it is hidden in the $H$ band due to a constant contribution from the  host galaxy and potentially from 
a foreground intervening galaxy.
In particular, the host galaxy is only the fourth GRB host at $z>6$ known to date.
By assuming a number density $n=1\,\mathrm{cm}^{-3}$ and an efficiency $\eta=0.2$, we derived a half-opening angle of $8.4^\circ\pm1.0^\circ$, which is the highest ever measured for a $z\gtrsim6$ burst, but within the range covered by closer events. The resulting collimation-corrected gamma-ray energy release of $\simeq1\times10^{52}$ erg is also among the highest ever measured.
The moderately large half-opening angle argues against recent claims of an inverse dependence of the half-opening angle on the redshift. The total jet energy is likely too large to be sustained by a standard magnetar, and it  suggests that the central engine of this burst was a newly formed black hole. 
Despite the outstanding energetics and luminosity of both GRB 210905A and its afterglow, we demonstrate that they are 
consistent within 2$\sigma$ with those of less distant bursts, indicating that the powering mechanisms and progenitors do not evolve significantly with redshift.
}


\maketitle

%
\begin{figure*}
\begin{center}
\includegraphics[width=\textwidth,angle=0]{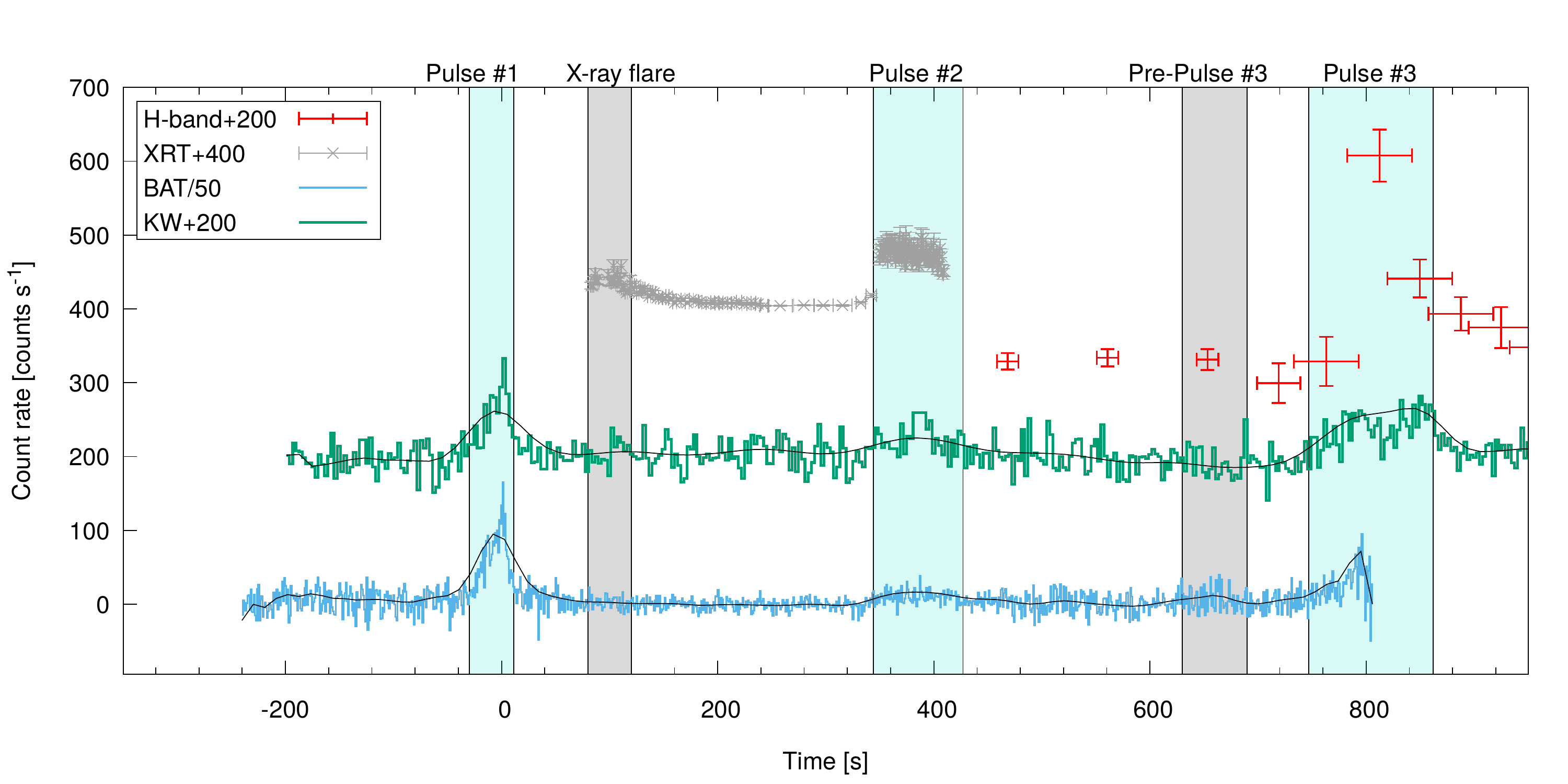}
\caption{Multi-band prompt emission light curve of  GRB~210905A. 
The light curve of GRB 210905A as seen by \textit{Swift}/BAT ($15-350$ keV, 6 s binning, $\textnormal{count rate}/50$, cyan), Konus-\textit{Wind} ($20-400$ keV, 5.888 s binning, $\textnormal{count rate}+200$, green points), \swift/XRT ($0.3-10\textnormal{ keV, count rate}+400$, grey points) and REM (flux density$+200$ in units of $0.1\,Jy$, red points). The evolution of the gamma-ray emission is highlighted with a black smoothed spline to guide the eye. The intervals corresponding to the three main and two smaller pulses are highlighted by turquoise- and grey-shaded areas, respectively. 
}
\label{fig:gammaXopt}%
\end{center}
\end{figure*}

\section{Introduction} \label{sec:intro}

The discovery of a $z>6$ gamma-ray burst (GRB) is a rare occurrence
 that, thanks to the extreme luminosity of these sources, offers a window into the infant Universe, which is otherwise difficult to observe.  
Long GRBs, with gamma-ray emission generally 
longer than 2~s \citep{Kouveliotou1993a},
originate from the explosions of very massive stars \citep{Hjorth2003b,Stanek2003ApJ,WoosleyBloom2006a}.
Under the assumptions that the stellar initial mass function (IMF) in distant galaxies is not broadly different from that of closer objects and 
that the opening angles do not evolve strongly with redshift,
the rate of GRBs can be used both to estimate the star-formation rate (SFR) \citep{Kistler2009a,Robertson2012a} and to study the effects of metallicity on supernovae (SNe)-Ibc and GRB progenitors 
\citep{Grieco2012a}.  
The SFR is expected to change at very high redshift with the transition from the first massive population III (pop-III) stars in the remote Universe to pop-II and pop-I stars \citep{Salvaterra2015a,Fryer2022a}. 
How this happens remains an open question that may be addressed through GRB studies.
We note that the prompt emission is not affected by dust extinction, and thus GRBs can provide a census of obscured star formation at all redshifts \citep{Blain2000a}. 
Due to their immense brightness, GRBs can also act as beacons illuminating the local circumburst medium
\citep[e.g.][]{Savaglio2003a,Prochaska2008a,Schady2011a,Watson2013a,Heintz2018e}, 
the interstellar medium (ISM) of their hosts
\citep[e.g.][]{Fynbo2006a,Savaglio2012a,Cucchiara2015a,Bolmer2019a,Heintz2019a}, and the surrounding intergalactic medium (IGM) in the line of sight 
\citep{Totani2006a,Hartoog2015a}. They are therefore powerful probes of the ionisation and chemical enrichment history of the early universe.
To shed light on these open issues through very high-redshift GRBs, several mission concepts have been studied and proposed 
\citep[e.g.][]{Amati2018a,Tanvir2021a,White2021a}.
 
So far, out of the $\approx555$ GRBs with a  well-constrained spectroscopic redshift (as of 20 July 2022), only five have been detected\footnote{See \url{http://www.mpe.mpg.de/~jcg/grbgen.html}} at $z\gtrsim6$: 
GRB 050904 \citep[$z=6.295$,][]{Kawai2006a,Tagliaferri2005a}, 
GRB 080913 \citep[$z=6.733$,][]{Greiner2009a,Patel2010AA}, 
GRB 090423A \citep[$z=8.23$,][]{Tanvir2009Nature,Salvaterra2009a},
GRB 130606A \citep[$z=5.913$,][]{Hartoog2015a,Chornock2013a}\footnote{We consider this burst to be at $z\sim6$ since it lies just below this threshold.}, and GRB 140515A \citep[$z=6.327$,][]{Chornock2014arXiv,Melandri2015a}. 
An additional four have very low signal-to-noise spectra or photometric redshifts:
GRB 090429B \citep[$z\simeq9.4$,][]{Cucchiara2011a},
GRB 100905A \citep[$z\simeq7.88$,][]{Bolmer2018a},
GRB 120521C \citep[$z\simeq6$,][]{Laskar2014a}, and
GRB 120923A \citep[$z\simeq7.8$,][]{Tanvir2018a}.
Some of these events show larger prompt energetics than those at low redshift, but this is likely the result of observational biases, and a cosmic evolution of the GRB energy release function has not been confirmed yet \citep[e.g.][and references therein]{Tsvetkova2017,Tsvetkova2021}. 
In fact, also very-high redshift GRBs follow the $E_{\mathrm{peak,}z}-E_\mathrm{iso}$ and $E_{\mathrm{peak,}z}-L_\mathrm{iso}$ correlations 
\citep[`Amati' and `Yonetoku' correlations; ][]{Amati2002a,Yonetoku2004ApJ}.
The same is true for the afterglow luminosity (Kann et al. 2022a, in prep.), which is larger only when compared with the low-luminosity local events ($z<0.2$).
The large prompt energy release is well matched by a larger X-ray luminosity of their afterglows, as indeed the $L_{\mathrm{X}}/E_{\mathrm{iso}}$ is similar to that of low-redshift events. These results suggest that the powering mechanisms and progenitors do not evolve with redshift.
On the other hand, some studies have suggested that jets from GRBs in the high-redshift universe are more narrowly collimated than those at lower redshifts \citep[e.g.][]{Lloyd-Ronning2019a,Laskar2014a,Laskar2018a}.

Here we present a follow-up of the bright GRB 210905A, the tenth burst with redshift $z\gtrsim6$ detected in the last 16 years. 
It was detected by the
\textit{Neil Gehrels Swift Observatory} ~\citep[][\textit{Swift} hereafter]{Gehrels2004a} and \kw ~\citep{Aptekar1995a}. X-ray as well as optical and near-infrared (NIR) follow-up observations of its bright afterglow led us to determine a spectroscopic redshift of $z=6.312$ \citep[refined with respect to][]{Tanvir2021GCN30771}.
 The burst was also detected by the Cadmium Zinc Telluride Imager (CZTI) on-board {\it Astrosat} \citep{Prasad2021GCN30782} and, following the detection by the \swift{} Burst Alert Telescope (BAT, \citealt{Barthelmy2005SSRv}), it was also found via a targeted search in data of the Gamma-ray Burst Monitor (GBM) on-board {\it Fermi} \citep{Veres2021GCN30779}. 

In \S\ref{sec:data} we describe the observations of both the GRB and the afterglow, and in \S\ref{sec:mod} we present the analysis of the data. In \S\ref{sec:dis} we discuss the results and compare them with other bursts at low and high redshift, and we draw our conclusions in \S\ref{sec:con}.
Throughout this work, the flux density of the afterglow is described as $F_\nu (t) \propto t^{-\alpha} \nu^{-\beta}$.  A $\Lambda$CDM cosmological model with $\Omega_M = 0.308$, $\Omega_{\Lambda} = 0.692$, and $H_0 = 67.8$ km s$^{-1}$ Mpc$^{-1}$ \citep{Planck2016a} has been assumed for calculations. 
All data are in the observer frame and $1\sigma$ errors are used throughout the paper, unless otherwise specified.

\section{Observations}\label{sec:data}

\subsection{Gamma-ray and X-ray observations.}
\label{sec:kw}

GRB 210905A was discovered by BAT on-board  \swift{} at $T_0=00$:12:41.3 UT on 5 September 2021 \citep{Sonbas2021GCN30765}. 
 The BAT light curve shows a complex
structure with three pulses, detected until  $\sim800$ s after the burst trigger. 

Since GRB 210905A was too weak to trigger\footnote{See \S4.3 of \cite{Tsvetkova2021} for details on the KW trigger sensitivity.} \kw{} (KW), 
the burst data are available only from the instrument's waiting mode, as first reported by \cite{Frederiks2021GCN30780}.
In this mode, count rates with a coarse time resolution of 2.944~s are recorded continuously 
in three energy bands: G1 ($20-100$~keV), G2 ($100-400$~keV), and G3 ($400-1500$~keV).
A bayesian block analysis of the KW waiting mode data in S1 (one of the two NaI(Tl) detectors)
reveals three (separated in time) emission episodes, 
each featuring a statistically significant count rate increase in the combined G1+G2 band (Figure~\ref{fig:gammaXopt}),
while no statistically significant emission was detected in the G3 band throughout the burst. 

The first episode, which triggered \swift/BAT,
started at $\simeq T_0-30$~s and ends at $\simeq T_0+11$~s (hereafter Pulse~1).
The weaker second episode ($\sim T_0+344$~s to $\sim T_0+426$~s; Pulse~2) coincided in time with the bright flare in the XRT windowed-timing (WT) mode light curve around $T_0+400$~s (Figure~\ref{fig:gammaXopt}).
The onset of the final emission episode, observed by KW from $\sim T_0+747$~s to $\sim T_0+862$~s (Pulse~3), 
is clearly visible in the BAT mask-weighted data, which are available up to $\sim$800~s after the trigger.
The $T_\mathrm{90}$ duration\footnote{The total duration ($T_\mathrm{100}$) derived from the KW observation is $\sim$890~s (at the $5\sigma$ level).} of the GRB~210905A prompt emission derived from the KW observation is $\sim870$~s. 

\swift/XRT started observing the BAT error circle $91.7$~s after the trigger and found an unknown X-ray source at the UVOT-enhanced position coordinates RA (J2000) = 20$^{\rm h}$36$^{\rm m}$11$\fs$64, Dec. (J2000) = $-$44$^{\circ}$26\arcmin24\farcs3 with a final uncertainty of 1\farcs5 \citep[][\swift/XRT catalogue]{Beardmore2021GCN30768}. 
Pointed \swift{} observations continued until $3.8$ Ms after the GRB, when the source became too faint to be detected. Light curves and spectra, as well as the result of their modelling, have been obtained from the \swift/XRT repository \citep{Evans2007a,Evans2009a}.
 However, to build more accurate multi-wavelength spectral energy distributions (SEDs), given that some data available in the \swift/XRT repository suffer from bad centroid determination, we have processed the \swift{} data corresponding to the epochs of our SED analysis (obs. IDs 01071993001/002/003, Fig. \ref{fig:sedoptx}). 
 To reduce the data, the software package \texttt{HeaSoft} 6.29 was used\footnote{
\url{http://heasarc.gsfc.nasa.gov/docs/software/lheasoft}} with
the latest calibration file available\footnote{\swift/XRT calibration files: 20210915.}. For the data processing, we used standard procedures, consisting of the use of the package {\tt xrtpipeline}, available within the {\sc FTOOLS} distribution\footnote{\url{http://heasarc.gsfc.nasa.gov/ftools/}}, with standard-grade filtering. 
 Using the most refined position provided by the \swift{} team, the selection of the GRB position in the X-ray data
and the extraction of both source and background spectra, were done with the {\tt xselect} package, while for the construction of the corresponding ancillary response file (.arf) we used {\tt xrtmkarf} on each corresponding epoch exposure file. 
 In the following, a Galactic equivalent hydrogen column density of $N_H=3.38\times10^{20}\, \textnormal{cm}^{-2}$ is adopted \citep{Willingale2013a}.

\input{texdata.tex}

\subsection{Optical/NIR imaging and photometry}
\label{sec:optnir}

\swift/UVOT started observing about 156 s after the trigger but no credible afterglow candidate was found \citep{Siegel2021GCN30785}. The MASTER Global Robotic Net \citep{Lipunov2010a} was also pointed at GRB 210905A  6 s after notice time and 414 s after trigger time but could not detect any afterglow candidate \citep{Lipunov2021GCN30766}.

We obtained optical/NIR observations with the $0.6$m robotic Rapid Eye Mount telescope \citep[REM,][]{Zerbi2001a}, starting 428~s after the burst.
A transient source was detected immediately in the $H$ band and later in $i^\prime z^\prime ZJK$ bands (i.e. all except $g^\prime$ and $r^\prime$). 
 Observations continued for about 3 hr before the declining afterglow brightness fell below the instrument detection limits in all filters \citep{Davanzo2021GCN30772}. Images were automatically reduced using the jitter script of the \texttt{eclipse} package \citep{Devillard1997a} which  aligns and stacks the images to obtain one average image for each sequence. A combination of IRAF \citep{Tody1993} and SExtractor packages \citep{bertin2010sextractor} were then used to perform aperture photometry. 

We triggered Bessel $R$- and $I$-band observations with the 1m telescope of the Las Cumbres Observatory Global Telescope (LCOGT) network, equipped with the Sinistro instrument, at the Cerro Tololo Inter-American Observatory (CTIO), Chile. 
The midpoints of the first epoch are $t_I=1.06$ hr and $t_R=1.29$ hr, in the $I$ and $R$ bands respectively. 
The data provided by the LCO are reduced using the BANZAI pipeline \citep{mccully2018real} that performs bias and dark subtraction, flat-fielding, bad-pixel masking, and astrometric calibration. 
Afterwards, we use our own pipeline, which aligns and stacks the images using the astroalign Python package \citep{beroiz2020astroalign}, and afterwards uses SExtractor to perform the photometry and calibration against a sample of USNO-B catalogue stars \citep{monet2003usno}.
Using the data-reduction pipeline from LCO, and our relative photometry pipeline\footnote{The photometry was confirmed after the cross-calibration mentioned below.}, we calculate a magnitude of $I=19.46\pm0.15$ mag and a $3\sigma$ upper limit of $R>22.44$ mag. The lack of an $R$-band detection alerted us to the possibility that this burst may lie at very high redshift \citep[$z>5$, first reported by ][]{Strausbaugh2021GCN30769,Strausbaugh2021GCN30770}.

GRB 210905A was observed simultaneously in $g^\prime r^\prime i^\prime z^\prime JHK$ with
GROND \citep[Gamma-Ray Burst Optical Near-Infrared Detector;][]{Greiner2008a,Greiner2019PASP} 
mounted on the 2.2m MPG telescope at ESO La Silla Observatory in Chile \citep{Nicuesa2021GCN30781}. The first epoch observations were done around 23 hr after the GRB trigger. The afterglow was detected only in the $z^\prime JHK$ bands. A second set of observations obtained 7 hr later was shallower and yielded only upper limits. Subsequent follow-up observations were obtained on 7 and 8 September 2021, but the afterglow was also not detected in the latter epochs.
We continued our ground-based follow-up using both the \textit{VLT}/HAWK-I \citep[High Acuity Widefield K-band Imager,][]{Pirard2004a} NIR imager on Paranal, as well as the  
Dark Energy Camera (DECam) mounted on the 4m Victor Blanco telescope at CTIO. 
We also used the acquisition camera of the ESO \textit{VLT}/X-shooter spectrograph to obtain $g^\prime r^\prime I_{\rm Bessel} z^\prime $ imaging before moving on to spectroscopy.
We obtained a last ground-based observation 87 d after the GRB with \textit{VLT}/FORS2 in the $I_{\rm Bessel}$ band.

Finally, the field was observed with the \textit{Hubble Space Telescope (HST)} on 24 April 2022. At this epoch four dithered observations with a total duration of 4797 s were obtained in the $F140W$ filter. The data were obtained from the MAST archive and processed with {\tt astrodrizzle} to create a final combined 
charge transfer efficiency corrected 
image with a pixel scale of 0\farcs07/pixel.
Aperture photometry was performed with a radius of 0\farcs4  to minimise any contribution from the nearby sources (see Figure \ref{fig:forshawk}).

X-shooter and GROND optical/NIR images were reduced in a standard manner using PyRAF/IRAF \citep{Tody1993}. In particular, GROND data reduction  was done with a customised pipeline \citep{Kruhler2008a} that is based on standard routines in IRAF. 
FORS $I$-band and HAWK-I $JHK_s$-band data have been reduced using the ESO Reflex environment \citep{esoreflex2013a}. 
We obtained PSF photometry with the DAOPHOT and ALLSTAR tasks of IRAF. PSF-fitting was used to measure the magnitudes of the GRB afterglow. 
Only for the late-time FORS2 observation in $I_{\rm Bessel}$ at 87 days and \textit{HST}-$F140W$ at 232 days did we use aperture photometry. 

All optical photometry except $I_{\rm Bessel}$-band data were calibrated against the SkyMapper catalogue \citep{Wolf2018a}, while the ground-based NIR photometric calibration was performed against the 2MASS catalogue \citep{Skrutskie2006a}. This procedure results in a typical systematic accuracy of 0.04~mag in $g^\prime r^\prime i^\prime z^\prime$, 0.06~mag in  $JH$ and 0.08 mag in $K_s$. 
To cross-calibrate all the $I$-band imaging we
applied the Lupton formulae to a set of local standard stars from the SkyMapper catalogue.

The $I$ filters used by X-shooter and LCO extend beyond 10000~{\AA}. Therefore, we expect that not all the flux is dimmed by the Lyman-$\alpha$ dropout at $\sim8900$~{\AA} in these filters. On the contrary, the FORS2 $I$-band filter has negligible transmission above Lyman-$\alpha$ (at the redshift of GRB 210905A). Therefore, we speculate that the possible (note the large error) late $I$-band emission (see \S~\ref{sec:constant}) does not originate from the afterglow, but instead from a foreground source.

The optical/NIR afterglow lies at coordinates RA (J2000) = $20^h36^m11\fs57$, Dec. (J2000) = $-44^{\circ}26\arcmin24\farcs7$ 
as measured in our first HAWK-I image and calibrated against field stars in the GAIA DR2 catalogue \citep{Gaia2018a} with the astrometric precision being 0\farcs15. This refines the position 
reported by LCO \citep{Strausbaugh2021GCN30769} and is in agreement with the more 
precise localisation provided by ALMA \citep{Laskar2021GCN30783}.
Table~\ref{tab:photall} provides a summary of all photometry of the transient (non-relevant upper limits are not reported). All reported magnitudes are in the AB photometric system and not corrected for the Galactic foreground extinction of $E(B-V)=0.029$ mag \citep{SchlaflyFinkbeiner2011a}.


\subsection{X-shooter spectroscopy and redshift}
Starting $\sim2.53$ hr after the GRB detection, we obtained UV to NIR spectroscopy 
of the afterglow with the X-shooter instrument \citep{Vernet2011a} mounted on the \textit{VLT} on Cerro Paranal (ESO, Chile), via the Stargate Large Programme for GRB studies.

The afterglow is well detected in the red part of the visible arm. 
A clear break is detected around 9000~{\AA}, which we interpret as the Lyman-$\alpha$ break (first reported in \citealt{Tanvir2021GCN30771}). Other lines such as 
\ion{Fe}{ii}, \ion{Al}{ii}, \ion{C}{iv} and \ion{Si}{ii}
and fine structure lines are visible and display two velocity components, which belong to the ISM of the same galaxy. All these lines allow us to determine $z=6.312$ as the redshift of the GRB. 
A very strong foreground system at $z = 2.8296$ (\ion{Mg}{ii}, \ion{Fe}{ii} lines) 
and another at $z = 5.7390$ (\ion{C}{ii}, \ion{Fe}{ii}, \ion{C}{iv}, \ion{Si}{iv} lines) are also present.
The details concerning the reduction and analysis of the absorption lines in the X-shooter spectra are given in \cite{Saccardi2022a}.
This high redshift explains the non-detection by UVOT and MASTER and the red $R_C-I_C$ and $r^\prime-z^\prime$ colours found with LCO and X-shooter as due to Lyman dropout. In \cite{Fausey2022a} we will study the IGM neutral fraction in light of the GRB 210905A afterglow spectrum.


\begin{figure}
\begin{center}
\includegraphics[width=\columnwidth,angle=0]{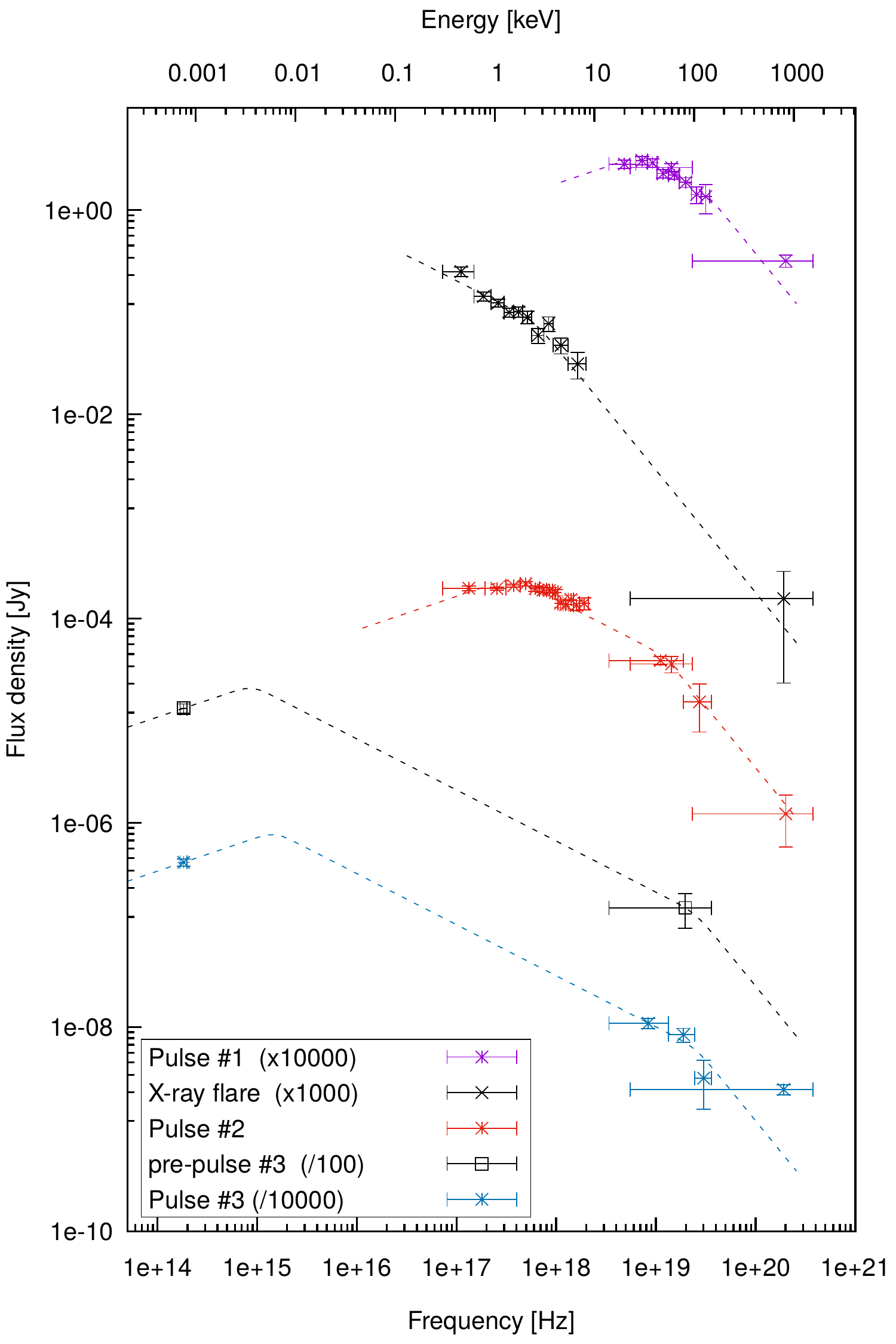}
\caption{Optical/NIR to gamma-ray SEDs of the prompt phase at five different epochs (see \S~\ref{sec:gammaxo}). All SEDs have been modelled with a double broken power-law following the expectations from synchrotron theory. Note that we could not constrain the low-energy break during the X-ray flare. In the fourth SED, we have simply scaled the solution from the last epoch (there is no KW detection during this epoch).  Note that the photon indices described in the text correspond to spectral indices $1/3$, $-1/2$, $-1.3$ shown here. X-ray data are corrected for Galactic and intrinsic absorption.}
\label{fig:sedgammaxo}%
\end{center}
\end{figure}

\begin{table*}
	\centering   
\footnotesize
	\caption{Fits to the prompt emission spectra.}
	\label{tabFits}
\setlength{\tabcolsep}{0.6em}
\begin{threeparttable} 
	\begin{tabular}{lcccccccc}
	\toprule
		Spectrum$^a$ & Instruments &Model$^{f}$ &Time interval          & $\alpha$	& $E_\textrm{peak}$ or $\nu_m$ &  $E_\textrm{break}$ or $\nu_c$         & Flux (15--1500 keV)        &$\chi ^{2} (\textrm{d.o.f.})$	\\
			&    &     & (relative to $T_0$, s) & (photon index)	& (keV)		        &   (keV)	 &$(10^{-7}$ erg cm$^{-2}$ s$^{-1})$ &\\
	\midrule
		`peak'$^b$& BAT+KW&CPL  & [$-0.465$, $2.479$]     &$-0.66_{-0.30}^{+0.35}$ & $144_{-28}^{+56}$    & & $2.83_{-0.40}^{+0.56}$ &40.2 (58)\\[2ex]
	\midrule \\
		Pulse~1& BAT+KW &CPL & [$-29.905$, $11.311$]   & $-0.99_{-0.17}^{+0.18}$ & $127_{-19}^{+31}$   & & $1.14_{-0.11}^{+0.14}$  &36.5 (58)\\
		
		Pulse~1& BAT+KW& DBPL  & [$-29.905$, $11.311$]  &   & 127   &  $27.09_{-3.34}^{+3.41}$ &  & 35.2 (56)\\[4ex]
		
		X-ray flare$^{c}$& XRT+KW& DBPL  & [$80.0$, $120.0$]  &   & $1.46_{-0.26}^{+0.21}$  & unconstrained &  & 225.9 (281)\\[4ex]
		
		Pulse~2$^{c}$& XRT+BAT+KW& CPL& [$343.983$, $426.415$]  & $-0.80_{-0.47}^{+0.58}$ & $70_{-13}^{+22}$      & & $0.28_{-0.05}^{+0.06}$&48.6 (58)\\	
		Pulse~2$^{c}$& XRT+BAT+KW&DBPL & [$343.983$, $426.415$]  &   & 50 & $1.13_{-0.10}^{+0.11}$  &    & 714.2 (714)\\[4ex]
		
		Pulse~3$^{d}$& BAT&CPL & [$747.311$, $797.359$] & $-0.89_{-0.30}^{+0.37}$ & $154_{-36}^{+77}$    &  & $0.76_{-0.11}^{+0.17}$& 55.2 (58)\\ 	
		Pulse~3$^{e}$& KW &CPL & [$747.311$, $862.127$] & $-0.88_{-0.30}^{+0.76}$ & $167_{-61}^{+88}$    & & $0.97_{-0.23}^{+0.24}$ & 0 (0)$^{e}$\\
		Pulse~3$^{d,g}$& REM+BAT+KW&DBPL & [$747.311$, $797.359$]  &   & 154 & $0.006$\tnote{g} & & 59.3 (56)\\	
	\bottomrule
	\end{tabular}
\begin{tablenotes}
		\item[a]{All spectra, except the first, are time-averaged.} 	
		\item[b]{This spectrum was used to calculate the peak energy flux.} 	
		\item[c]{The interval covered by XRT.} 
		\item[d]{The interval covered by BAT.} 
		\item[e]{KW-only fit; for the CPL model, the 3-channel fit has 0 degrees of freedom.}  
		\item[f]{CPL stands for cut-off power-law. DBPL stands for double-broken power-law, used for the synchrotron model. In this last case, the power-law indices were fixed as described in \S\ref{sec:gammaxo}.}
		\item[g]{This break was fixed to match the $H$-band data.
		}
\end{tablenotes}
\end{threeparttable}    
\end{table*}


\section{Modelling and results}\label{sec:mod}

\subsection{Joint BAT-KW modelling}\label{sec:kwbat}

To derive the broad-band spectral parameters of the prompt emission of this burst, we performed a joint spectral analysis of the BAT data ($15-150$ keV) and the KW waiting-mode data ($20-1500$~keV)
for all three prompt emission episodes in a way similar to that described in \cite{Tsvetkova2021}. 

The spectral data from the two instruments were simultaneously fit in
\texttt{Xspec v12.12.0} using three different spectral models (see below), all normalised to the energy flux in the $15-1500$ keV range.
The most reliable results for all three emission episodes were obtained with a power-law function with high-energy exponential cutoff (CPL). 
Compared to the CPL, a simple power-law (PL) function fits the data with significantly worse statistics ($\Delta \chi^2 > 7$ in all cases) 
and systematically overestimates the high-energy part of the spectra.
The Band function \citep{Band1993a} does not improve the fit statistics as compared to the CPL. 
For all spectra, the Band fits\footnote{The Band function has parameters 
$\alpha$, $\beta$ and $E_\mathrm{peak}$, not to be confused with the decay and spectral indexes of the afterglow, defined in \S\ref{sec:intro}.} provide values of the index $\alpha$ and $E_\mathrm{peak}$ almost identical to the CPL fits (and consistent within the large errors), 
and set only an upper limit to the high-energy photon index ($\beta < -2.3$),
due to the sparse KW data which do not provide enough sensitivity and spectral resolution to constrain the spectral index above 100 keV.

Our fits with the CPL function are summarised in Table~\ref{tabFits}.
The time-averaged spectrum of the brightest episode (Pulse~1) is best described by $\alpha \sim -0.99$ and observed $E_\mathrm{peak} \sim$127~keV. 
The spectrum of the weaker episode (Pulse~2) is characterised by a similar, within errors, $\alpha$, and an about halved $E_\mathrm{peak} \sim$70~keV. 
The third emission episode is  $\sim 115$~s long and
only partially covered by \swift/BAT. In this case, we analysed the spectra extracted for two time intervals: 
the first spectrum corresponds to the time interval of joint KW and BAT detection ($\alpha \sim -0.89$, $E_\mathrm{peak} \sim$154~keV), 
and the second one covers the whole third emission episode ($\alpha \sim -0.88$, $E_\mathrm{peak} \sim$167~keV). 
For the latter interval, the fits were made using the KW 3-channel spectrum alone and the obtained model flux was used to calculate the Pulse~3 energy fluence.

The $15-1500$ keV energy fluences of Pulses~1, 2, and 3, derived from our time-averaged fits, are summarised in Table \ref{tabEiso}, together with the
fluence integrated over all three emission episodes. We use these results to calculate the isotropic energy (see also \S\ref{sec:con-prompt}).
The spectrum in the interval ($T_0-0.465$ s, $T_0+2.479$ s) inside Pulse~1, which corresponds to the peak count rate, 
is characterised by $\alpha \sim -0.66$ and $E_\mathrm{peak} \sim$144~keV.
Using this spectrum and the BAT light curve, we estimate the 1~s peak energy flux of GRB~210905A to be $3.83_{-0.54}^{+0.73} \times 10^{-7}$~erg~cm$^{-2}$~s$^{-1}$ ($15-1500$ keV).


\subsection{Joint modelling of the prompt emission from gamma-rays to the optical}
\label{sec:gammaxo}

In the previous section we have analysed the gamma-ray spectra during the three pulses and found that they can be modelled 
almost equally well with a CPL or a Band function with very similar (within errors) low-energy photon index $\sim-0.8<\alpha<-1.2$ and $E_{\rm peak}$.
The high-energy index of the Band function is $\beta<-2.3$, poorly constrained by the sparse KW data.
Values of $-1$ and $-2.3$ are very typical low- and high-energy photon indices for GRBs \citep[e.g.][]{Preece1998a,Nava2011a}. Following early works \citep{Frontera2000a,Rossi2011a,Zheng2012a}, recently \cite{Oganesyan2019a}
have shown that the low-energy spectra ($<100$ keV) of the majority of \swift/BAT GRBs actually have
a low-energy spectral break in the $2-30$ keV range, in addition to the typical break
corresponding to the peak energy at larger energies. 
Such a break has also been discovered at higher energies, up to few hundreds of keV, in \textit{Fermi} bursts \citep{Ravasio2018a,Ravasio2019a}, and has been studied in detail \citep{Gompertz2022} in the temporally long merger event GRB 211211A \citep{Rastinejad2022}. It has been suggested to be a common feature of GRB prompt emission spectra \citep{Toffano2021a}.
Therefore, the low-energy part of the spectrum, with photon index $-1$, 
into two power-law photon indices describing the spectrum below and above the
low-energy break, and have distributions centred around $-2/3$ and $-3/2$ (or $1/3$ and $-1/2$ for the flux density spectrum $F_\nu$), respectively.
These indices are the same as those 
below and above the cooling break $\nu_c$
and expected by the synchrotron theory in the fast-cooling regime \citep[see also][]{Ravasio2018a,Ravasio2019a}.
Further confirmation of these empirical fits was obtained by direct fitting of prompt GRB spectra with a synchrotron model \citep{Ronchi2020a,Burgess2020a} and the synchrotron interpretation is discussed for example in  \cite{Ghisellini2020a}.

To determine if the prompt emission of GRB 210905A is in agreement with these theoretical expectations,
we have modelled the NIR and X- to gamma-ray SEDs of five epochs during the whole prompt emission with a double broken power-law with photon indices fixed to  the synchrotron model predictions.
That the optical-to-gamma emission is the result of a common radiative process is justified by the simultaneous evolution of the optical, X-ray and gamma-ray prompt emission.
The selected epochs are the three gamma-ray pulses, the first X-ray flare at $\sim120$~s and 
an additional epoch at $\sim630-690$ s simultaneous to an $H$-band observation. This is the only 
epoch before the last pulse with few counts in the BAT spectrum. We have fixed the high-energy break ($\nu_m$, the frequency corresponding to the minimum injection energy in a fast-cooling synchrotron model) to the break energy in the Band modelling above. 
The high-energy photon index above this break has been fixed to $-2.4$, that is also consistent with the Band fit. 
The results are shown in Table \ref{tabFits} and in Figure \ref{fig:sedgammaxo}. 
The analysis of the X-ray flare alone shows that it is well modelled by $\nu_m$ at $\sim1$ keV, a photon index\footnote{We could not constrain the low-energy break for this epoch.} $-2.4$
and intrinsic absorption $N_H=7.7^{+3.6}_{-3.2}\times10^{22}\,  \textnormal{cm}^{-2}$. 
In the following, we fixed the intrinsic hydrogen column density to this value.
During the first two pulses the data are consistent with a broken power-law with photon index 0.5.

In the last two SEDs, we also include the $H$-band follow-up obtained with REM (Figure~\ref{fig:gammaXopt}). 
Note that in the fourth SED we have simply scaled the solution from the last epoch, because there are basically just two measurements for three possible free parameters\footnote{Two breaks and the peak flux.}, not enough to constrain all breaks. Therefore, this is not shown in Table~\ref{tabFits}.

For these last SEDs (i.e. before and during Pulse 3) the $H$-band observation is below the extrapolation of the photon index from the gamma-rays, and thus $\nu_c$ must be in between the $H$ and X-ray bands. We further discuss the implications of this finding in \S\ref{sec:ori-prompt}.
Unfortunately, for both SEDs the lack of any colour information  and possible contribution from the emerging afterglow 
in the observed optical/NIR prevents us from affirming without doubt that the low-energy photon index is $-2/3$. However, we can confirm that for both SEDs the synchrotron model is in agreement with the observations. 

\begin{figure*}
\begin{center}
\includegraphics[width=0.99\textwidth,angle=0]{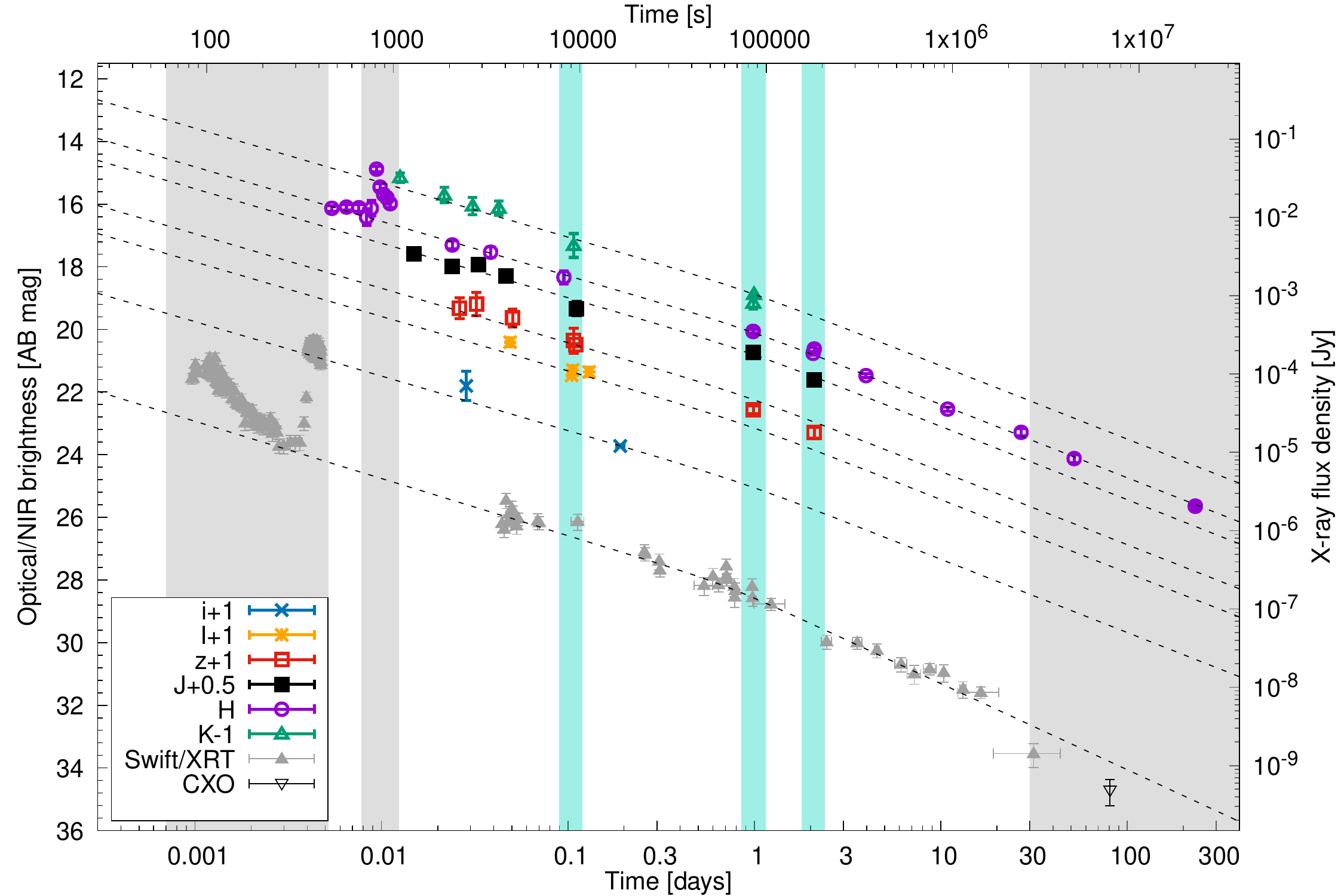}
\caption{Optical and X-ray light curves. The dashed lines show the fit to each single band assuming a smoothly broken power-law model. The grey intervals are not considered in the first modelling of the light curves. Those in light blue have been used for the SED fitting (see \S \ref{sec:xo}). The X-ray light curve is computed at 1.73 keV, the log-mean of the XRT band. The last $H$-band data corresponds to the \textit{HST}/F140W detection. No  colour correction was necessary as explained in  \S\ref{sec:break}.}
\label{fig:lcoptx}%
\end{center}
\end{figure*}

\begin{figure}
\begin{center}
\includegraphics[width=\columnwidth,angle=0]{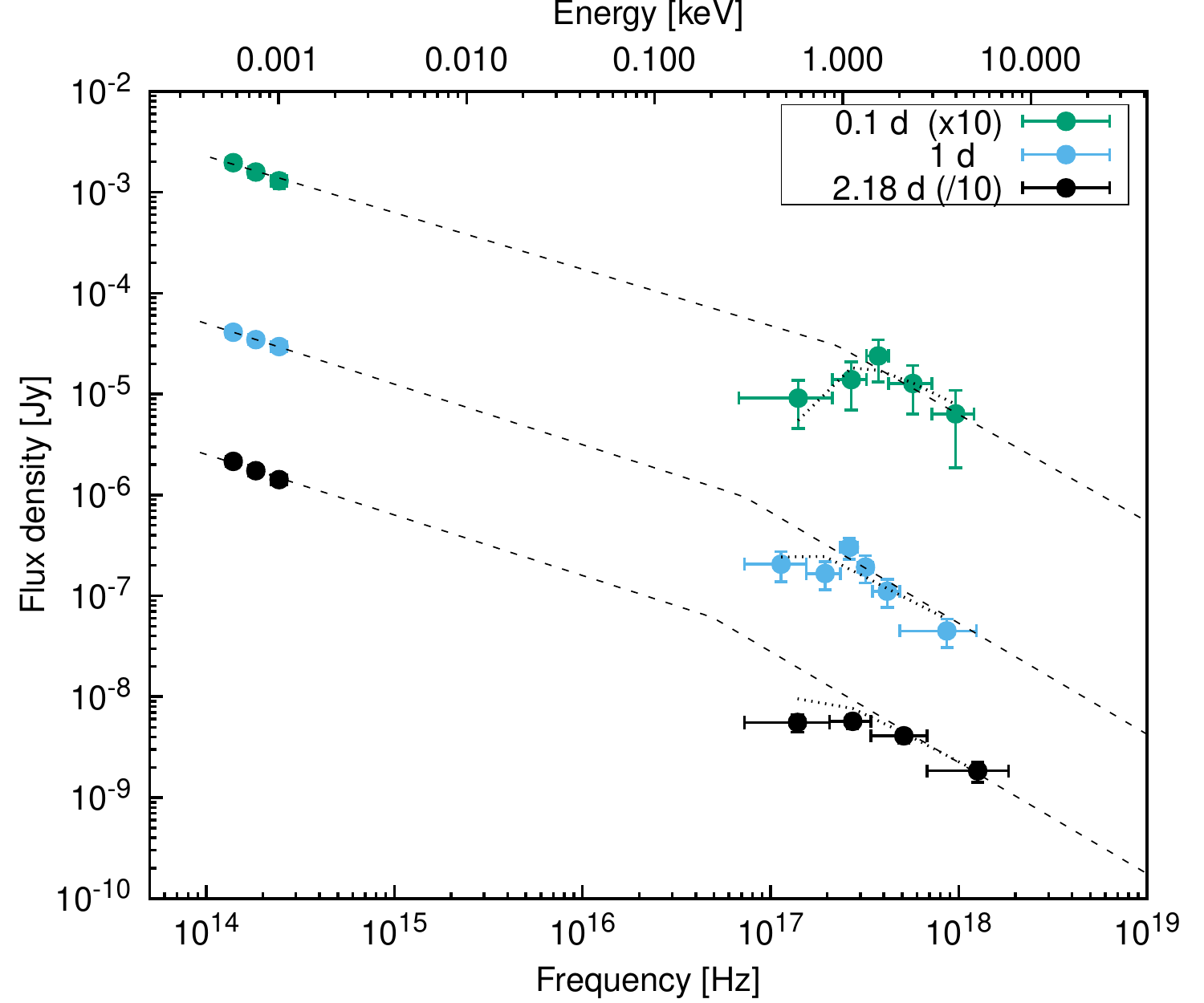}
\caption{Optical/NIR to X-ray SEDs of GRB 210905A at three different epochs (0.1, 1.0, 2.18 d). The best fit with a broken power-law is shown in all three epochs, and the best-fit parameters are shown in Table \ref{tab:sed} (see \S \ref{sec:xo}). The dotted- and dashed-lines show the absorbed and unabsorbed models, respectively.
}
\label{fig:sedoptx}%
\end{center}
\end{figure}

\begin{table}
\centering
\caption{Optical/NIR to X-ray modelling of the afterglow (see Figure~\ref{fig:sedoptx}).}
\begin{threeparttable}
\setlength{\tabcolsep}{0.3em}
\begin{tabular}{lccccc} 
\toprule
Model &Time & $\beta_\mathrm{opt}$\tnote{a} & \multicolumn{2}{c}{Break (keV)} &    $\chi^2/$ d.o.f. \\ 
&d     &                 & obs.        & Theor.\tnote{b}      &            \\ 
\midrule  
BPL&  0.1   &     $0.62\pm0.04$ &  $1.7_{-1.6}^{+2.6}$  & $1.4\pm1.2$  &    9.8/16  \\  
SPL&  0.1   &     $0.63\pm0.03$ &  --  & --  &   9.9/17  \\
BPL&  1.0   &     $0.60\pm0.04$ &  $0.35_{-0.13}^{+0.28}$  & 0.35  &   63.1/53  \\ 
SPL&  1.0   &     $0.71\pm0.02$ &  --  & --  &   75.8/54  \\
BPL&  2.2  &     $0.56\pm0.16$ &  $0.18_{-0.03}^{+0.06}$ &  $0.22\pm0.19$  &    22.8/25       \\  
SPL&  2.2   &     $0.80\pm0.03$ &  --  & --  &   19.5/26  \\ 
\bottomrule                                 
\end{tabular}   
\begin{tablenotes}
\footnotesize 
\item[a] In the broken power-law we assumed $\beta_{X}=\beta_\mathrm{opt}+0.5$. 
\item[b] Obtained from the best-fit value at 1 d with  $\nu(t)=\nu(1d)(t/1d)^{-k}$, with $k=0.5$ after 1 d (ISM, slow cooling scenario) and $k=0.6$ before 1 d assuming energy injection (see \S\ref{sec:eninj}).  
\end{tablenotes}    
\label{tab:sed}
\end{threeparttable}         
\end{table}


\subsection{Joint afterglow light curve and SED}\label{sec:xo}

Figure \ref{fig:lcoptx} shows both optical/NIR and X-ray light curves of the afterglow. Regions in grey have not been considered in this section because of: i) the presence of flares likely due to long-lasting activity from the central engine
and ii) a possible late break when compared to the earlier evolution (Figure~\ref{fig:lcoptx}) that we discuss below in \S \ref{sec:break}.

A complete understanding of the afterglow behaviour would require a full numerical simulation. Nevertheless, we can derive some conclusions by modelling the SEDs and light curves of the afterglow.
We modelled the afterglow SED from NIR to X-ray frequencies at three different epochs, 0.1, 1.0, and 2.18 days, using \texttt{Xspec v12.12.0} \citep{Arnaud1996a}. We have not considered the optical data ($z$-band and bluer bands) because they are affected by the Lyman-$\alpha$ break and thus do not add anything useful to this modelling.
The redshift was fixed to 6.312 and we fixed the Galactic and intrinsic hydrogen column density (see \S\ref{sec:kw}).
To avoid being affected by the uncertain gas absorption, we have not considered data below 0.5 keV in the modelling. 
We have modelled the NIR-to-X-ray SED 
both with a single and a broken power-law with $\beta_{X}=\beta_{\rm opt}+0.5$, at all three epochs. The best-fits
are shown in Figure~\ref{fig:sedoptx} and their
parameters are shown in Table~\ref{tab:sed}.
All fits give negligible dust extinction $A_V\lesssim0.03$ mag, independent of the extinction
law\footnote{In the \texttt{zdust} model.}, which is not unusual for high-$z$ GRB afterglows (see \S\ref{sec:avnh}).
It is not straightforward to decide between the single and  broken power-law models as the SEDs are fit comparably well in both cases. However, we note that in the first epoch they give basically the same value for the low-energy spectral index.  Therefore, we conclude that $\nu_c$ is within or above the X-ray band at $0.1$~d. That $\nu_c$ is then in between the two bands is even more clear in the second SED at 1~d whose best-fit gives $\beta_\mathrm{opt}=0.60\pm0.04$, and thus an electron index $p=2.20\pm0.08$. To confirm these findings, we need to also consider the light-curve evolution. 

We have modelled the optical and NIR light curves simultaneously with a smoothly broken power-law \citep{Beuermann1999a}: 
$F = (F_1^{\kappa}+ F_2^{\kappa})^{-1/\kappa}$,
where $F_\textrm{x}=f_\textrm{break}(t/t_\textrm{break})^{-\alpha_x}$,
$f_\textrm{break}$ being the flux density at break time $t_\textrm{break}$, $\kappa$ the break smoothness parameter, and the subscripts $1,2$ indicate pre- and post-break, respectively. We find a shallow break with large uncertainty at $t_\textrm{break,~opt} =0.99\pm 0.73$ d ($85.9 \pm62.7$~ks) 
and  decay indices $\alpha_{1,\rm opt}=0.69\pm0.04$ and $\alpha_{2,\rm opt}= 0.94 \pm 0.04$, with break smoothness $\kappa=10$ fixed \citep{Zeh2006a}\footnote{We have also evaluated smaller fixed $\kappa$ values (5, 2, 1) and find that $\chi^2/\mathrm{d.o.f.}$ increases, 
$t_b$ remains similar, but even at $\kappa=5$, the error exceeds the value of the break time, and increases further.}.
With respect to a simple power-law, the $\chi^2/\mathrm{d.o.f.}$ decreases from $1.36$ to $0.92$.  

The X-ray light curve shows an initial  
peak at $97$~s followed by the typical steep decay \citep{Tagliaferri2005Nature,Barthelmy2005ApJ} with $\alpha=2.37^{+0.15}_{-0.16}$ until $\sim270$ s after the burst, when it is interrupted by a flare also visible in gamma-ray data. 
After $\sim3000$ s, it is best modelled by a broken power-law with a 
shallow break at $\simeq 1$~d: from $\alpha_{1,X} = 0.74^{+0.03}_{-0.01}$ to $\alpha_{2,X} = 1.10\pm 0.04$, with $t_\textrm{break,~X}=60\pm30$~ks\footnote{As shown by the \swift/XRT light curve repository \citep{Evans2007a,Evans2009a}.} (1$\sigma$ errors). 
Finally, we note that modelling simultaneously the X-ray and optical bands with the best-fit indices found above, the shallow break is seen at a common time of $0.70\pm0.26$ days ($60.5\pm22.5$ ks).

In Table \ref{tab:closure}, we compare the observed evolution with the predicted values of the temporal slopes in the optical/NIR and the X-ray bands for various slow-cooling afterglow scenarios \citep[see, e.g. ][]{Zhang2006c,Schulze2011a} and the electron index $p=2.20\pm0.08$.  We cannot find a good solution for the data before $0.7$ d, 
however, after the first modest break  
the data are best modelled within a scenario where the jet is expanding into a constant-density medium 
(hereafter referred to as the interstellar medium
or ISM environment). A single power-law SED solution  
cannot explain the observed temporal decay index in X-rays, $\alpha_{2,X}=1.1$, with emission below the cooling frequency, $\nu_c$. Moreover,  within this solution $\beta_\mathrm{opt}$ should be constant, but instead it evolves with time.
These results indicate that $\nu_c$ should lie between the optical and X-ray bands (see also Figure \ref{fig:sedoptx}). 
A $\nu_c$ that has moved out of the X-ray band can explain the difference in the temporal decay index between optical and X-rays after the shallow break, therefore, we consider a broken power-law as the best description for the optical-to-X-ray SED.
For an upper branch\footnote{The spectral index $\beta_\mathrm{X}=0.90\pm0.15$ at $22.8$~ks reported in the XRT pages is well in agreement with this result.} $\beta_{X}=1.10\pm0.04$, obtained at 1d (the epoch with the best statistics), the electron index is $p=2.20\pm0.08$. 
The large errors on the cooling frequency do not permit to test whether the break shifts in time as $t^{-1/2}$, although the results seem consistent with such a relation (see Table \ref{tab:sed}).

\begin{table}
\centering
\caption{Closure relations.}
\centering
\small
\setlength{\tabcolsep}{0.3em}
\begin{threeparttable}
\begin{tabular}{lccc}
\toprule
Afterglow model	& Theoretical & \multicolumn{2}{c}{Observed} \\
		& $\alpha$  &   $\alpha_\textrm{1,opt}=0.69\pm0.04$		& $\alpha_\textrm{1,X}=0.74\pm0.03$ \\
		&           &$\sigma$-level\tnote{a}&  $\sigma$-level\tnote{a}\\
\midrule
ISM\tnote{c}, wind, $\nu>\nu_c$		&$1.15\pm0.06	$	&$6.38$	&   $6.11$	  	\\
ISM, $\nu<\nu_c$		&$0.90\pm0.06    $  &$2.88$&   $2.36$   	\\
wind, $\nu<\nu_c$		&$1.40\pm0.06	$	&$-9.76$	&    	$-11.03$	   \\
\midrule
			        &  &   $\alpha_\textrm{2,opt}=0.93\pm0.04$	& $\alpha_\textrm{2,X}=0.93\pm0.04$	\\
\midrule
ISM, wind, $\nu>\nu_c$		&$1.15\pm0.06	  $    &$3.02$& 0.69\tnote{b} \\
ISM, $\nu<\nu_c$		&$0.90\pm0.06$      &$-$0.50\tnote{b}&  $-2.77$\\
wind, $\nu<\nu_c$		&$1.40\pm0.06	$	&$-13.17$&    	$-15.25$	   \\
\bottomrule
\end{tabular}
\begin{tablenotes}
\footnotesize     
\item[a] The $\sigma$-level is the difference of the predicted and
the observed temporal slope, normalised to the square root of the sum of their quadratic errors. 
\item[b] The solution that matches the closure relations within 1 $\sigma$ is highlighted in bold (see \S\ref{sec:xo}).
\item[c] We follow the common use and refer to the constant-density medium as ISM.
\end{tablenotes}
\label{tab:closure}
\end{threeparttable}     
\end{table}


\begin{figure}[htp]
\centering
\includegraphics[width=0.95\columnwidth,angle=0]{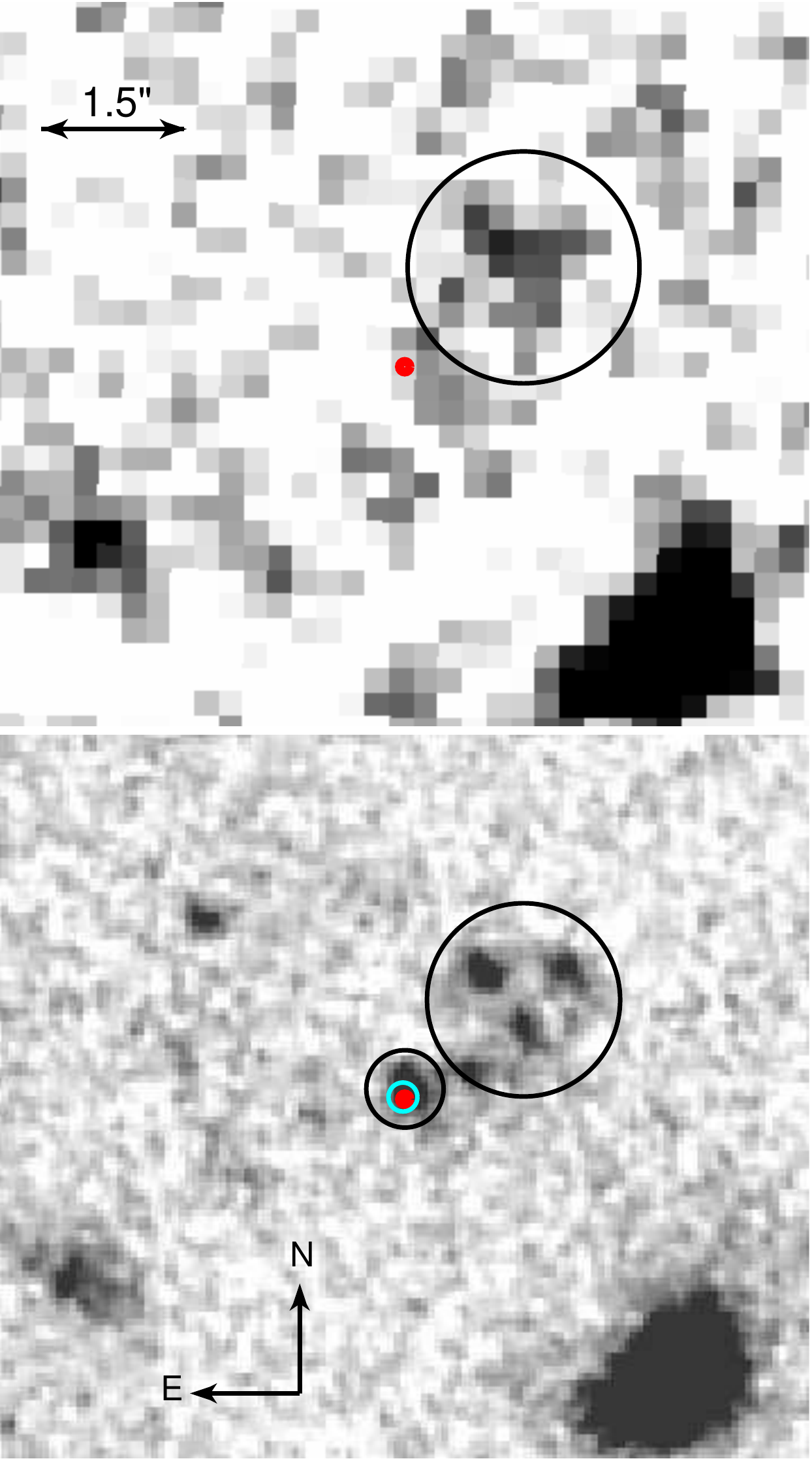}
\caption{Zoom-in to the $10\arcsec\times10\arcsec$ region centred on the afterglow. 
\textit{Top:} deep FORS2 $I_{\rm Bessel}$-band image obtained 87 d after the GRB trigger. The red point indicates the ALMA localisation of the afterglow. A faint source, highlighted with a black circle, lies 
$\sim1\farcs5$ NW of the afterglow position. 
The radius of the circle is that of the aperture used for photometry. 
\textit{Bottom:} For comparison, the \textit{HST}/$F140W$ image shows several sources 
at the position of the $I_{\rm Bessel}$-band source.  
The cyan circle shows the location of the NIR afterglow and its error (\S\ref{sec:optnir}), measured in the first HAWK-I $H$-band observation.
}
\label{fig:forshawk}
\end{figure}

\subsection{The late NIR imaging}\label{sec:hst}

In Figure \ref{fig:forshawk} we show 
the most recent observation of the field
obtained with \textit{HST} in the $F140W$ band. 
At 0\farcs09$\pm0\farcs02$ from the NIR afterglow position we clearly detect an extended source ($F140W=25.66\pm0.05$ mag). The relative offset is measured comparing the centroids in the first HAWK-I image and the \textit{HST} image, after aligning these two images using a common set of sources. 
It is slightly elongated in the NNE-SSW direction and has a FWHM of 0\farcs4 and 0\farcs3, larger than the FWHM of field stars ($0\farcs25\pm0\farcs02$). 
Therefore, we conclude that the \textit{HST} detection is dominated by a constant source. 
The statistical probability of chance alignment is $P_{cc} (<r)=0.03$ \citep{Bloom2002a}, which has been obtained using the projected angular separation ($r=0\farcs4$), the apparent magnitude ($H_{AB}=25.8$ mag, see \S\ref{sec:break}), and the $H$-band galaxy counts from \cite{Frith2006a}.  
This is lower than  what is commonly used to establish an association. Therefore, it is likely that this is the host galaxy of the GRB. This source is not detected in the $I_{\rm Bessel}$-band in an observation 
obtained 87 d after the GRB with the \textit{VLT}/FORS2 instrument down to a $3\sigma$ upper limit of $26.0$ mag (AB). 
We also note a more complex structure  
which extends up to 2\farcs2 to the NW of the afterglow position. 
This extended structure is weakly detected in the deep $I_{\rm Bessel}$ observation  
with a similar brightness ($I_{\rm Bessel,AB}=24.84\pm0.18$ mag and $F140W_{AB}=24.7\pm0.04$ mag), and therefore is unlikely to reside at $z=6.3$.
This group of sources, or at least part of them,   
could also be responsible for the foreground intervening system found in X-shooter spectra at $z=2.8$ with high-EW \mbox{Mg\,{\sc ii}} absorption \citep[see][]{Saccardi2022a}. In this case, the cold gas observed in absorption can also be offset from the hot and bright region observed in the $I$-band, or occupy a larger region of the same foreground galaxy.

\subsection{Constraints on the jet break}\label{sec:break}

A sizeable number of GRB afterglow light curves
break to steeper power-law decays, usually within a few days after the trigger. 
These breaks have generally been interpreted as due to the outflow being collimated in a jet, where the break occurs when the relativistic beaming angle becomes wider than  
the jet's half-opening angle $\theta_{\rm jet}$  \citep{Rhoads1997a,Sari1999a}.
In the forward-shock model the jet breaks have to be achromatic, thus to have the same slope (and slope change) simultaneously in all bands\footnote{The value of the light-curve post-jet-break slope depends on $\nu_\mathrm{obs}$ being above or below $\nu_m$ and $\nu_{sa}$, where $\nu_{sa}$ is the synchrotron self-absorption frequency. Optical and X-ray afterglow SEDs are usually observed to be above $\nu_m$ \citep[e.g.][]{Greiner2011a}.}.

In \S\ref{sec:xo} we have shown that a moderate break is present in both optical and X-rays at a common time of $\simeq 0.74$ d. 
However, the post-break slope for both X-rays and optical is only $\simeq 1$, that is too shallow for a jet break,
both  observationally \citep{Wang2015a}
and theoretically \citep{SariPiran1999a,Zhang2006c,Panaitescu2007a}.  
Instead, the last XRT detection, together with the late observation by the \textit{Chandra} X-ray Observatory \citep[][]{Laskar2021GCN31127},
shows that the light curve breaks at $\sim30$ d.
However, the NIR light curve, taken up to  
232 d in the observer frame, shows no simultaneous steep break (Figure~\ref{fig:lcoptx}).
In the following, we assume that the break in X-rays is indicative of an achromatic break, and the last NIR detection is likely dominated by another component
(see \S\ref{sec:hst},\ref{sec:constant}). Note that we do not apply any  colour correction between $F140W$ and $H$ bands because the UV slope is basically flat for GRB host and star-forming galaxies  \citep[e.g.][]{Schulze2015a}\footnote{Using the spectral slope $\beta_\mathrm{opt}=0.6$ obtained from the SED fitting of the afterglow, the  colour correction is just $H-F140W=-0.10$ mag, and thus will not make an appreciable difference in our analysis.}. 

\begin{figure}
\centering
\includegraphics[width=\columnwidth,angle=0]{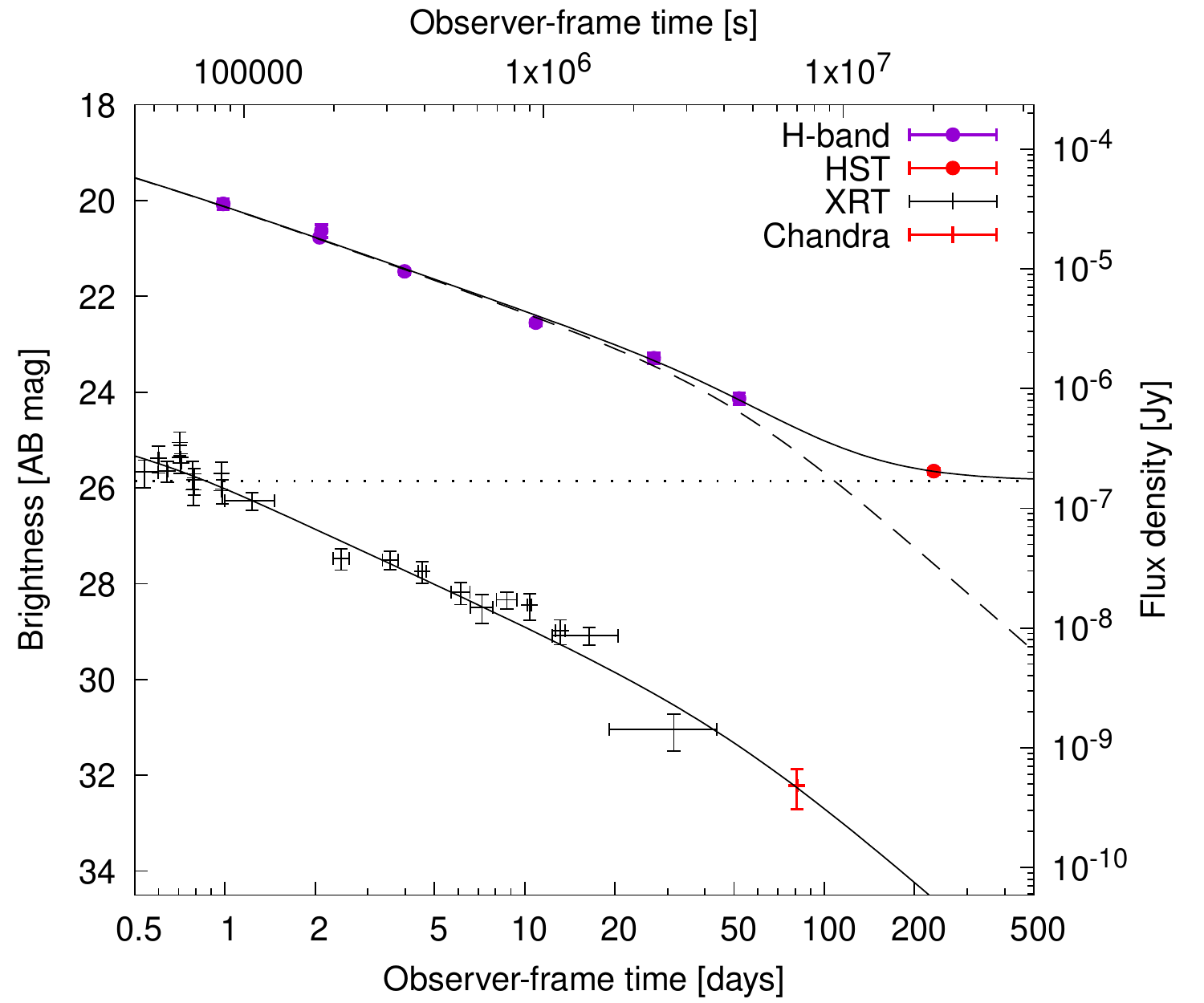}
\caption{Observer-frame $H$-band (purple) and X-ray (black) light curves. Solid lines show the joint fit with a smoothly-broken power-law, assuming common achromatic breaks. The dashed line shows the $H$-band light curve without constant component. 
The horizontal dotted line shows the modelled $H$-band constant component.
Following the slow-cooling scenario (\S\ref{sec:xo}), we fixed the pre-break decay indices to $\alpha_\mathrm{opt}=0.9$ and $\alpha_\mathrm{X}=1.15$. The last break is interpreted as a jet break, and thus the post-break decay index has been fixed to $\alpha_\mathrm{opt,X}=p=2.2$. The late flattening in the $H$-band light curve can be explained by a constant contribution from a host or intervening system. See \S\ref{sec:break} for details.
}
\label{fig:latelchx}
\end{figure}

To better constrain the break time, we modelled jointly the $H$-band and X-ray light curves after the early break at $0.7$ d with a smoothly broken power-law 
$F = (F_2^{\kappa}+ F_3^{\kappa})^{-1/\kappa}$, following the definition in \S\ref{sec:xo}  
but with the subscripts $2,3$ indicating the pre- and post-jet break respectively. We fixed the pre-jet-break index to the model values $\alpha_\mathrm{2,opt}=0.9$ and $\alpha_\mathrm{2,X}=1.15$ (see Table \ref{tab:closure}). 
In our analysis we adopt the jet model (with sideways expansion) and slow cooling \citep[e.g.][]{Sari1998a,Zhang2004a}.
Therefore, we assume that the post jet-break index is $\alpha_\mathrm{3,opt}=\alpha_\mathrm{3,X} \simeq p=2.2$ (see \S\ref{sec:xo}).
Note that the sparse data after the break prevent us from constraining the $\kappa$ parameter. Therefore, we 
let it vary between the two extremes of the interval $1<\kappa<5$. These are consistent with the expected values for emission either side of $\nu_c$ and for a typical GRB observation angle \citep{vanEerten2013a, Lamb2021a}. We note that from the models, it is difficult to get $\kappa>5$ \citep{vanEerten2013a}.
In the $H$-band we have considered an additional constant component (see \S \ref{sec:constant}).
In summary, the only free parameters are the break time, the flux at the jet break and the flux of the constant source.
The best-fit break time in the observer frame is\footnote{Assuming no sideways expansion, and thus $\alpha_3=\alpha_2+3/4$ \citep{Panaitescu2007a}, we find a similar solution $t_{\rm jet}=36.4\pm21.6$\,d and thus similar half opening angle and conclusions.} 
$t_{\rm jet}=46.2\pm16.3$\,d
with the constant source having $H_{AB}= 25.8\pm0.2$ mag.
The modelling is shown in Figure \ref{fig:latelchx}.

\section{Discussion}\label{sec:dis}

\begin{figure}
\begin{center}
\includegraphics[width=0.48\textwidth,angle=0]{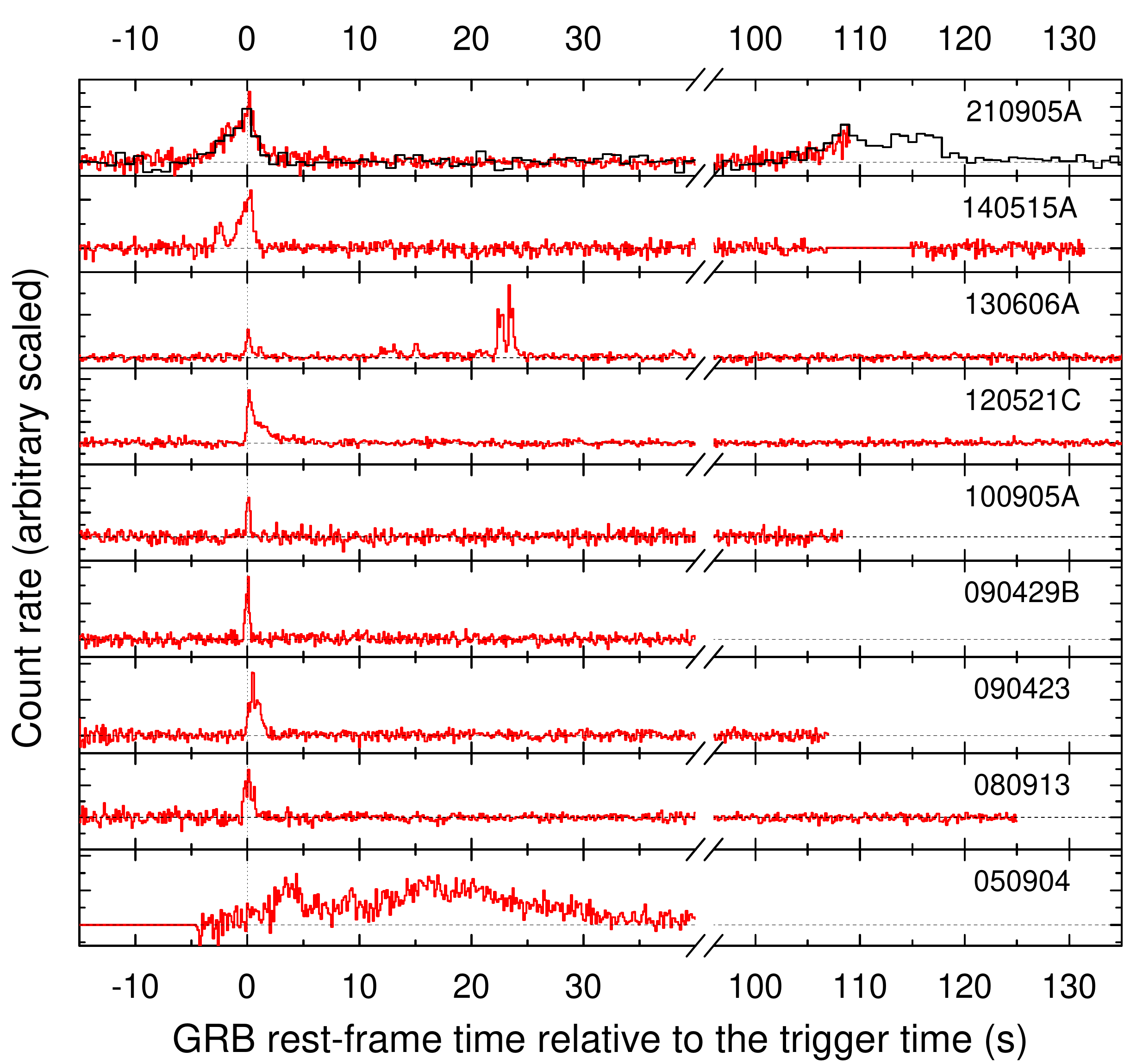}
\caption{GRB~210905A prompt-emission light curve (top panel) in the context of eight high-redshift GRBs ($z\geq6$). Arbitrarily scaled BAT count rates ($15-350$ keV) are plotted in red against time in the GRB rest frame. The KW light curve of GRB~210905A is plotted in black. The vertical dotted line shows the trigger time.
}
\label{fig:lcgammaz}%
\end{center}
\end{figure}

\begin{figure*}
\begin{center}
\includegraphics[width=0.89\textwidth,angle=0]{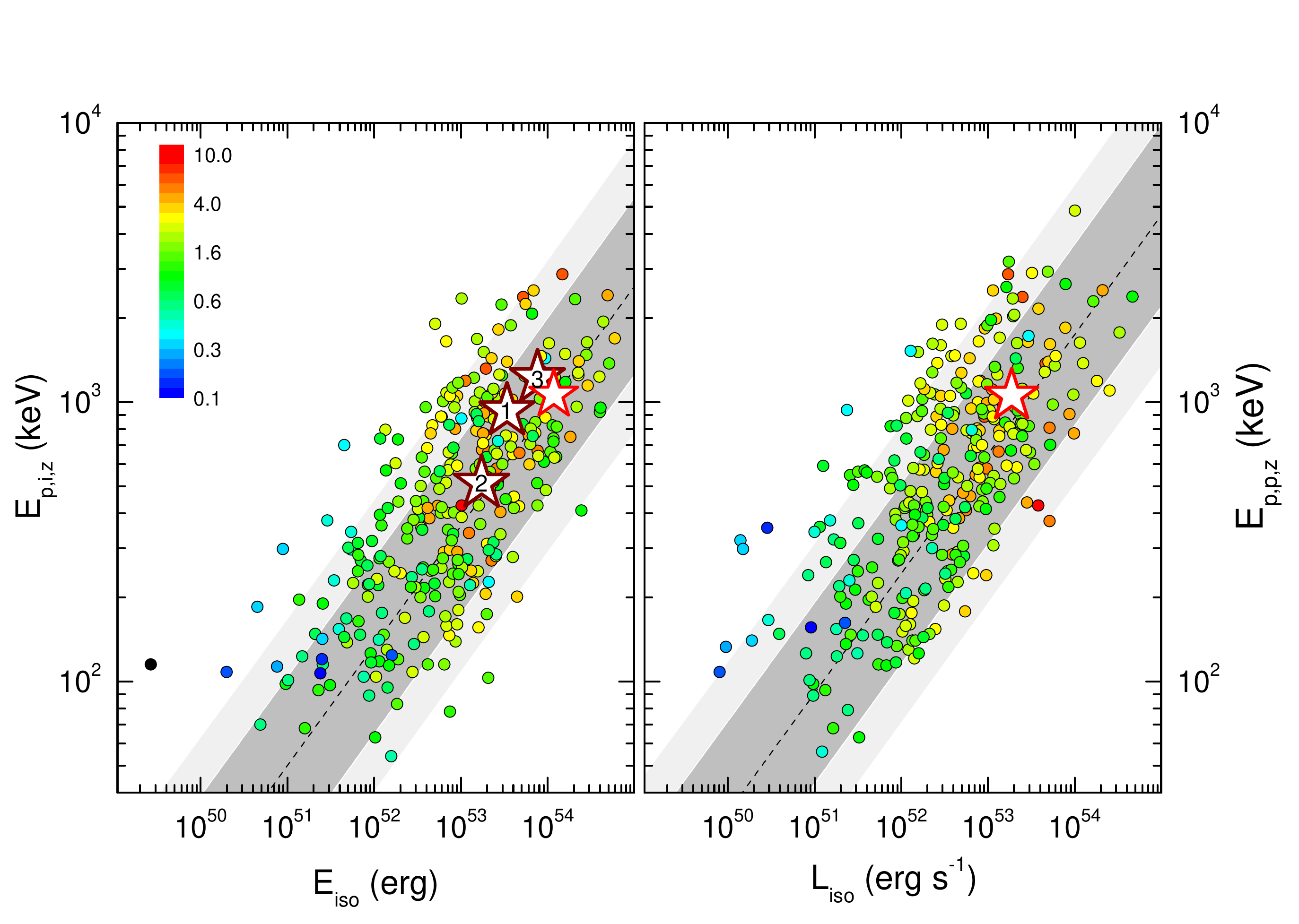}
\caption{Rest-frame energetics of GRB~210905A in the $E_{\mathrm{peak,}z}$--$E_\mathrm{iso}$ and $E_{\mathrm{peak,}z}$--$L_\mathrm{iso}$ planes (red stars).
		Brown stars in the left panel show the values derived for the individual Pulses 1, 2, and 3. 
		The rest-frame parameters of 315 long KW GRBs with known redshifts \citep{Tsvetkova2021} are shown with circles; the colour of each data point represents the burst's redshift.
		In the left plot, rest-frame peak energy values are derived from time-averaged spectral fits ($E_{\mathrm{p,i,}z}$).
		In the right plot, they are derived from spectra, corresponding to the burst's peak count rate ($E_{\mathrm{p,p,}z}$).
		The `Amati' and `Yonetoku' relations for this sample are plotted with dashed lines and the dark- and light-grey shaded areas show their 68\% and 90\% prediction intervals, respectively. The error bars are not shown here for reasons of clarity. 
}
\label{fig:amati}%
\end{center}
\end{figure*}

\subsection{The nature of the prompt emission}
\label{sec:ori-prompt}

GRB 210905A is among the few exceptional cases where optical data could be obtained during a gamma-ray pulse (Figure~\ref{fig:gammaXopt}). In the past, in less than a dozen cases
has modelling of the prompt emission been possible 
from optical/NIR to gamma-rays, such as in the cases of GRBs 990123, 041219A, 060526, 080319B, 080603A, 080928, 090727, 091024, 110205A, 111209A, 130427A, and the more recent GRBs 160625B and 180325A
\citep[e.g.][]{SariPiran1999a,Vestrand2005a,Thoene2010a,Racusin2008Nature,Guidorzi2011a,Rossi2011a,Kopac2013a,Virgili2013a,Stratta2013a,Kann2011a,Zheng2012a,Gendre2013a,Vestrand2014a,Troja2017Nature,Becerra2021a}.

At $z>6$, this analysis was possible only for GRB 050904  \citep{Boer2006a}.
In all these cases, modelling of the data with a broken power-law shows that the X-to-gamma-ray SED of the prompt pulses is in agreement with synchrotron emission, and in particular with fast cooling.
This is in agreement with studies on large samples
as we have mentioned in \S~\ref{sec:gammaxo}
\citep[see][ for a discussion on the possible implications]{Ghisellini2020a}.

However, when including the optical data the situation can be more complex: for example, the main and earlier pulses of GRBs 990123 \citep{SariPiran1999a,Galama1999a,Corsi2005a,Maiorano2005a}, 080319B \citep{Racusin2008Nature,Bloom2009ApJ}, 111205A \citep{Zheng2012a}, 130427A \citep{Vestrand2014a}, 160625B \citep{Troja2017Nature}, and 180325A \citep{Becerra2021a} show a convex spectrum between optical and X/gamma-rays.  Although different interpretations are also possible \citep[e.g.][]{Guiriec2016a}, this feature can be explained by synchrotron emission from internal forward shocks dominating the gamma-ray and X-ray prompt emission, while the early optical flashes are generated by a reverse shock.

The analysis of Pulse 3 of GRB 210905A is however clearly in disagreement with this latter scenario, with the $H$-band emission being fainter than the extrapolation of the power-law modelling the gamma-rays (Figure \ref{fig:sedgammaxo}).
Therefore, although simultaneous optical-to-gamma coverage of the first prompt pulses is missing in the case of GRB 210905A, we show that at least during the last pulse there is no indication that the NIR data have an origin  different from the X/gamma-ray emission, and all the observed epochs during the prompt phase can be explained by synchrotron emission from internal shocks.
This is not surprising, as in several events (e.g. GRBs 990123, 130427A, 160525B) the optical-to-gamma SED later evolves and can be entirely explained as emission from the forward shock. \cite{Oganesyan2019a} have shown that the later SEDs are consistent with being produced through synchrotron emission in the moderately fast-cooling regime from the same emission region.
 
\subsection{Prompt emission in context}
\label{sec:con-prompt}

Using $z=6.312$, we estimate the rest-frame properties 
of the burst prompt emission.  
Isotropic-equivalent energy release ($E_\mathrm{iso}$) and rest-frame spectral peak energies $E_{\mathrm{peak,}z} = (1+z) E_\mathrm{peak}$
for the individual emission episodes were calculated 
from the CPL spectral fits (\S\ref{sec:kwbat}); they are listed in Table~\ref{tabEiso}.
Integrated over the three intervals, the total energy release of GRB~210905A in $\gamma$-rays is $E_\mathrm{iso} = 1.27_{-0.19}^{+0.20} \times 10^{54}$~erg,
which is within the highest $\sim$7\% for the KW sample of 338 GRBs with known redshifts \citep{Tsvetkova2017,Tsvetkova2021}.
Since $E_\mathrm{peak}$ obtained from our fits differs between the individual emission episodes, 
we used the spectral peak energy value weighted by the episode fluence, $E_\mathrm{peak} \sim $ 145~keV, to estimate the burst time-averaged $E_{\mathrm{peak,}z}$ to $\sim1060$~keV.
This intrinsic peak energy is among the highest $\sim15\%$ of long KW GRBs.
Derived from the peak energy flux, the peak $\gamma$-ray luminosity of the burst is $L_\mathrm{iso} = 1.87_{-0.26}^{+0.36} \times 10^{53}$~erg~s$^{-1}$.
The rest-frame $E_\mathrm{peak}$ corresponding to the time interval around the peak luminosity is $\sim1050$~keV. 
The reported values of $E_\mathrm{iso}$ and $L_\mathrm{iso}$ were calculated in the rest frame 1 keV--10 MeV range.
All the quoted errors are at the $1\sigma$ confidence level.

With these estimates, GRB~210905A as well as its individual episodes lie inside the 68\% prediction interval (PI) of the 
$E_{\mathrm{peak,}z}-E_\mathrm{iso}$
(`Amati' relation; Figure~\ref{fig:amati}) 
for 315 long KW GRBs with known redshifts \citep{Tsvetkova2021}. 
Likewise, the burst peak luminosity and the corresponding $E_{\mathrm{peak,}z}$ perfectly fit the `Yonetoku' relation for the sample. 

Figure~\ref{fig:lcgammaz} shows the GRB 210905A prompt emission in the context of eight GRBs at $z\gtrsim6$.
With the rest-frame duration  
$T_\mathrm{90} /(1+z) \sim 119$~s
GRB~210905A is the intrinsically longest high-$z$ GRB detected to date and is also among the longest $\sim$3\% of bursts as compared to the whole KW catalogue\footnote{This sample does not include six KW ultra-long ($T_\mathrm{100} > 1000$~s) bursts, all at low-to-moderate redshifts $z\lesssim2$.}
\citep{Tsvetkova2017,Tsvetkova2021}, which covers the range $0.04\leq z \leq 9.4$. 
In this high-redshift sample, GRBs 210905A and 130606A are the only bursts with well-separated emission episodes. Except for this feature, they are similar to all other bursts which show short spikes with only moderate energy release. The exception is GRB 050904, which is similar to GRB 210905A in terms of energy released ($E_\mathrm{iso}=(1.33\pm0.14)\times10^{54}$ erg) but shows a $\sim30$ s long emission episode with two extended peaks (and at least a third episode observed in X-rays). 

The most powerful burst at low redshift is GRB 130427A at $z=0.3399$ \citep{Selsing2019a}. This GRB can be considered  as a good analogue of the energetic high-$z$ population because of its high energy release  \citep[][]{Perley2014a,dePasquale2016a}. 
Its prompt emission parameters are similar to those of Pulse 3 of GRB 210905A (see Table~\ref{tabEiso}): 
$E_{\mathrm{peak,}z}\sim1415$ keV and  
$E_\mathrm{iso}\sim9.4\times10^{53}$ erg \citep[][]{Tsvetkova2017}.
Accordingly, GRB 130427A and Pulse 3 lie very close
in the $E_{\mathrm{peak,}z}-E_\mathrm{iso}$ plane.
We should note, however, that the intrinsic durations of GRB 130427A and pulse 3 of GRB 210905A differ by factor of two ($T_{90,z}\sim10$ s for GRB 130427A versus $\sim19$ s for Pulse 3).
The initial light curve of GRB 130427A is somewhat similar to that of GRB 210905A since it starts with a large structured peak $\sim$20 s long  in the rest-frame, followed by a third peak starting at $\sim$100 s. This second pulse is, however, orders of magnitude weaker than the main pulse. So, GRB 130427A is not a `genuine' multi-episode GRB such as this work's burst or GRB 130606A, and is instead more similar to the other high-redshift GRBs.

\begin{table}
	\centering
\setlength{\tabcolsep}{0.4em}
	\caption{Parameters of the individual prompt emission pulses.}
	\label{tabEiso}
	
\begin{threeparttable}
	\begin{tabular}{lccc}
	\toprule
	Episode	& $E_{\textrm{peak,}z}$  & Fluence ($15-1500$ keV)\tnote{a} & $E_\textrm{iso}$\\
			& (keV)			&   $10^{-5}$~erg~cm$^{-2}$     & $(10^{53}$ erg)  \\
	\midrule
		Pulse~1 & $930_{-140}^{+230}$ &$0.471_{-0.046}^{+0.052}$ &$3.40_{-0.33}^{+0.41}$\\
		Pulse~2 & $510_{-95}^{+160}$   &$0.245_{-0.035}^{+0.050}$ &$1.73_{-0.33}^{+0.37}$\\		
		Pulse~3\tnote{b} & $1220_{-450}^{+640}$ &$1.11_{-0.26}^{+0.28}$ &$7.62_{-1.81}^{+1.89}$\\[2ex]
	\midrule
		Total\tnote{c}   & $1060_{-320}^{+470}$           &$1.82_{-0.28}^{+0.29}$ &$12.7_{-1.9}^{+2.0}$\\
	\bottomrule
	\end{tabular}
    \begin{tablenotes}
    \footnotesize 
	\item[a] Fluences were calculated using the fits with the CPL function from Table~\ref{tabFits}. 
	\item[b] Only the KW 3-channel spectrum is used.
	\item[c] This fluence is integrated over all three emission episodes.
	\end{tablenotes}
	\end{threeparttable}
\end{table}


\subsection{Collimation--corrected energy and central engine}
\label{sec:beam}

Knowing the value of the jet opening angle is
crucially important because it enables us to estimate the `true', collimation-corrected, energetics of the outflow \citep{Frail2001a,Ghirlanda2007a}.  
Numerical and analytical calculations \citep[e.g.][]{Sari1999a} have shown that the half-opening angle of the jet is related to the jet-break time. Following \cite{ZhangMacFadyen2009a} we calculate this angle  $\theta_{\rm jet}$ using the following equation for a uniform jet expanding in a constant-density medium:
\begin{equation}\label{angle}
    \frac{\theta_\text{jet}}{\text{rad}}
    =0.12~\left(\frac{E_{\rm kin,iso}}{10^{53}\,{\rm erg}}\right)^{-1/8}    \left(\frac{n}{\rm cm^{-3}}\right)^{1/8}~ 
    \left(\frac{t_{\rm jet}}{\rm day}\right)^{3/8}  (1+z)^{-3/8} \,,
\end{equation}

\noindent where $E_{\rm kin,iso}$ is the kinetic energy of the outflow assuming isotropy; $n=1\,\rm cm^{-3}$
is the number density of the medium, assumed to be constant; $t_{\rm jet}$ is the jet-break time (observer frame, see \S\ref{sec:break}), while $z=6.312$ is the redshift of the event. 
The kinetic energy is the one left after the prompt phase, and which later dissipates in the afterglow. Together with
the energy released as gamma-rays in the prompt phase\footnote{Here, $E_{\gamma,\rm iso}$ is the same as $E_{\rm iso}=12.7\times10^{53}$ erg of \S\ref{sec:con-prompt}.} $E_{\gamma,\rm iso}$, it represents part of the total GRB fireball energy 
$E_{\rm total,iso}=E_{\rm kin,iso}+E_{\gamma,\rm iso}$ \citep[e.g.][]{Zhang2004a,dePasquale2016a}.  
Assuming an efficiency\footnote{And thus $E_{\rm kin,iso}=(1/\eta -1)\,E_{\gamma,\rm iso}$.} $\eta=E_{\gamma,\rm iso}/E_{\rm total,iso}=0.2$ 
 we derive $\theta_{\rm jet}=0.147\pm0.017$~rad,
or $8.41\pm0.97$ degrees.
If we consider that the outflow is collimated, the
`true' gamma-ray energy of the jet is
$E_{\gamma}=E_{\rm \gamma,iso} ~(1-\cos(\theta_{\rm jet}))
\simeq 1\times10^{52}$~erg. 
The assumed efficiency is justified theoretically  \citep[e.g.][]{Guetta2001a} and by recent studies of GRB afterglows in the optical, X-rays and GeV gamma-rays  \citep[e.g.][]{Beniamini2015a}.
However, higher values are also possible, as suggested by some observations \citep{Zhang2007b,Lu2018a} and theoretical models \cite[e.g.][]{KobayashiSari2001a,ZhangYan2011a}.
As shown in \cite{dePasquale2016a}, the minimum $E_{\rm total}$ is obtained for $\eta = 3/4$. Lower efficiencies correspond to higher total energies. 
Therefore, with $\eta=0.2-0.75$ we can estimate the `total collimated energy' of the jet to be $E_{\rm total} \simeq E_{\rm \gamma}/\eta \simeq 3$--$8\times10^{52}$~erg.
We note that the dependence of $\theta_{\rm jet}$ on $n$ and the kinetic energy is rather weak (Eq. \ref{angle}). 
Thus, the total energy is not sizeably affected by the exact values of $n$ and $E_{\rm kin}$\footnote{Please note that assuming here a larger density would cause an even larger $\theta_{\rm jet}$ and $E_{\rm total}$ \citep{Cenko2011a,Granot-vanderHorst2014a}.}. 

The most widely discussed models  
of central engines of GRBs are accreting magnetars or accreting black holes.
 We can assume for a standard neutron star with mass $M\sim1.4\,M_\odot$ the maximum rotation energy to be in the range $3\times10^{52}$ erg \citep{LattimerPrakash2016a} -- $7\times10^{52}$ erg \citep{Haensel2009a}.
Therefore, our analysis allows us to disfavour a standard magnetar as central engine of this GRB. Only the most extreme magnetar models with $M\gtrsim2.1\,M_\odot$ and rotation energy $\sim10^{53}$ erg are not excluded \citep[see][]{Metzger2015a,Dallosso2018a,Stratta2018a}.  
 On the other hand, according to the Kerr metric \citep{Kerr1963a} the rotational energy $E_\textrm{rot}$ of a black hole can reach up to 29\%  of its total mass, which exceeds that of neutron stars by a full order of magnitude. Indeed, rotating black holes of mass $M\sim3\,M_\odot$ possess rotational energies up to $E_\textrm{rot}\sim 10^{54}$ erg \citep[e.g.][]{vanPuttenDellaValle2017a}. 
 Therefore, an energy budget of $\sim10^{53}$ erg can be conveniently extracted via the Blandford-\.{Z}najek mechanism \citep{BlandfordZnajek1977a}, thereby suggesting that the central engine of GRB 210905A may well be a rotating black hole.

\subsection{The early X-ray and optical/NIR afterglow}\label{sec:eninj}

As shown in \S\ref{sec:xo}, although the optical-to-X-ray SED at 0.1 d is in agreement with the cooling break lying within the X-ray band, both their light curves are not well explained by the standard fireball scenario before the common shallow break at $\sim0.7$ d.
 Here, we can investigate whether our data can justify the early decay and the  shallow break.  

First, the early break at $\sim0.7$ d is not well constrained but we can exclude that it is due to 
a wind-to-constant-density transition as the light-curve decline, in such a scenario, would become shallower and not steeper \citep[e.g.][]{Panaitescu2007a,Schulze2011a}. 
The times and the slopes instead make it an example of a `canonical' GRB X-ray afterglow light curve \citep{Nousek2006a,Zhang2007a}. 
Studying the canonical light curve, \cite{Zhang2007a} interpreted the break between the shallow segment with $\alpha\simeq0.7$ to the more `normal' segment with $\alpha\simeq1$ as the end of an `energy injection' phase.
During energy injection, the ejecta is still receiving energy, either from a long-lived central engine, or by  
slower ejecta shells that catch up 
with the leading shell. 
In other words, the mild break should be interpreted as cessation of energy injection. 
Following the relations in \cite{Zhang2006c}, where $q$ the energy injection index,
 we have (for ISM and $p=2.2$): $\alpha_{\rm opt} = ((2p-6)+(p+3)q)/4$, from which follows $q=0.84$ and $\alpha=0.69$ for $\nu<\nu_c$; for X-rays we obtain $\alpha_{\rm X} = ((2p-4)+(p+2)q)/4$, so $q=0.84$, and $\alpha=0.98$ for $\nu>\nu_c$. 
Using a stratified shell model with ejected mass
$M(>\gamma) \propto \gamma^{-s}$, where $\gamma$ is the Lorentz factor of the shell \citep[][]{Rees1998a}
and the relation between $s$ and $q$ parameters \citep[$s=(10-7q)/(2+q)$,][]{Zhang2006c}, we find that a value of $s\sim1.45$ fits the pre-break behaviour.
Equally, a magnetar central engine model that continuously injects energy as $L(t) \propto t^{-q}$ \citep[e.g.][]{DaiLu1998a}, can model the early decay with $q=0.84$.
Therefore, we cannot discard one model over the other, specifically stratified shell versus magnetar. However, as discussed in \S\ref{sec:break}, the energy constraints
likely limit the viability of a new-born magnetar as the power source of the energy injection.

We note also that the theoretical energy-injected $\alpha_{X,1}\sim0.98$ is larger than the value  observed (0.74). However, the theoretical value assumes $\nu>\nu_c$ but in Table \ref{tab:sed} we see that  $\nu_c$ is well within the X-ray band in the first day after the burst trigger.
The energy injection changes the way the cooling frequency evolves, that is $\nu_c \propto t^{(q-2)/2}=t^{-0.58}$ for $q=0.84$, and thus the cooling frequency evolves slightly faster whilst energy injection is happening: if the cooling frequency is at $\sim2$ keV at 0.1 d, then at 1 d it would have been at $0.5$ keV, and consistent with what we observe. 
Moreover, one should also consider that the $\nu_c$ break is likely smooth and covers a relatively large interval \citep[e.g.][]{GranotSari2002a}.
Therefore, the observed temporal decay index may well be somewhere between the values predicted for the $\nu<\nu_c$ and the $\nu>\nu_c$ cases, i.e, $0.69<\alpha_{X,1}<0.98$, in agreement with the observed value before $\sim$70 ks.


\subsection{The nature of the $H$-band flattening at late times}
\label{sec:constant}

The likely discovery of the host of GRB 210905A is a rare discovery, given that up to July 2022 only three hosts (those of GRBs 050904, 130606A, and 140515A) had been confirmed at $z>6$ \citep{McGuire2016a}, and four if we consider the 
possible detection of the GRB 060522 host \citep{Tanvir2012a}.

The observed brightness of the source detected with \textit{HST} in the $F140W$ band corresponds to a rest-frame $m_{1900\AA}\sim-21$ mag, which is consistent with the characteristic magnitude at 1600 {\AA} of $z=6-7$ galaxies \citep[e.g.][]{Bouwens2021a}. Therefore, such a galaxy is not unusual, although it is more luminous in the UV than galaxies that contribute the most to the star formation at these redshifts. In the following, we make use of the brightness of $H_{AB}=25.8$ mag resulting from the light curve fitting. 
A host galaxy at $z=6.3$
with such a brightness
and thus a rest-frame UV luminosity of $L_{\nu}=1.47\times10^{29}\,\mathrm{erg\,s}^{-1}\,\mathrm{Hz}^{-1}$, would have a SFR $\sim16\,M_\odot\, \mathrm{yr}^{-1}$ using equation 1 in \cite{Kennicutt1998a}.
This is certainly an acceptable value \citep[see also the discussion in][]{Saccardi2022a}, and in fact \cite{McGuire2016a} find that the $z\gtrsim6$ GRB hosts known to date likely have similar SFR, assuming a short-lived burst of star formation \citep[see also][]{Tanvir2012a}.
If its brightness is confirmed with further observations, the host of 210905A would also be the brightest. We caution however, that at this stage is not possible to separate some contamination from a possible foreground source
discussed in \cite{Saccardi2022a}, and thus the host can be fainter and the inferred SFR lower.

One could also speculate whether a SN can contribute to the final observation. However, note that a SN should reach an absolute magnitude of $m_\mathrm{2200}\sim-21$ mag in the far UV ($H$-band in the observer frame). 
This is four times more then the most luminous GRB-SN confirmed spectroscopically, SN2011kl associated with GRB 111209A \citep[$m_\mathrm{2735}\sim-19.6$ mag at peak,][]{Greiner2015Nature,Kann2019AA}, although \cite{Kann2021a} have recently claimed the existence of an even more luminous SN associated with GRB 140506A with $M_{g^\prime}\approx-20.5$ mag. 
As we find no evidence that GRB 210905A is more than just a very energetic but otherwise typical long GRB, there is no reason to claim the GRB would be accompanied by an extremely UV-luminous SN of a type not seen associated with GRBs before.

In the following we explore possible alternatives to the above interpretation of a jet break well visible in X-rays but hidden in the NIR by a constant source that becomes dominant. Thus, we consider the possibility that the afterglow still contributes substantially and the $H$ band can be modelled with a single power-law, with a chromatic break in the X-rays. It is also possible to speculate that the late light curve is the consequence of a spectral break moving between optical and X-rays. However, this not only contradicted by the elongated, extended, and offset nature of the \textit{HST} detection but, as we have shown in Section \S\ref{sec:xo}, our SED analysis which shows $\nu_c$ being already between the optical and X-ray bands after $0.7$ d. 
Therefore, it is not possible to invoke the presence of an additional break in the slow-cooling regime moving into the band after this time. Moreover, the change in the temporal index is inconsistent with the passage of the cooling break, which should be $\Delta\alpha=0.25$ in the slow-cooling and uniform-medium environment, and additionally incompatible with other regimes, for example fast cooling or a wind-blown environment.

Another possibility is to consider a bright reverse shock, but this requires either a large energy gradient or a big difference in shell velocity, both of which are inconsistent with the gradual energy injection scenario which explains the early light curve until 0.7 d. 
We could invoke a second, discrete shell with energy that is less than or comparable to the first (post initial energy injection) shell but much faster i.e. a delayed launch. However, the shell would have to conveniently collide with the leading shell at about the jet-break time and the optical excess would be the contribution from the reverse shock. This is not only an incredulous coincidence, but would also approximately double the total energy requirements to explain a second shell, making this event even more extreme.
We cannot exclude that a more mild shock could however explain the X-ray data at $10-20$ d, which lies just above the analytical modelling of the light curve.
We also cannot, however, confirm this possibility with the few data points available,
 which are anyway within $2\sigma$ of the analytical model.

In conclusion, 
we consider the detection of the host and/or an intervening galaxy (or a mix of the two) as the strongest and most plausible explanation for the flattening of the $H$-band light curve.

\subsection{The X-ray afterglow in context}

\begin{figure}
\centering
\includegraphics[width=1\columnwidth]{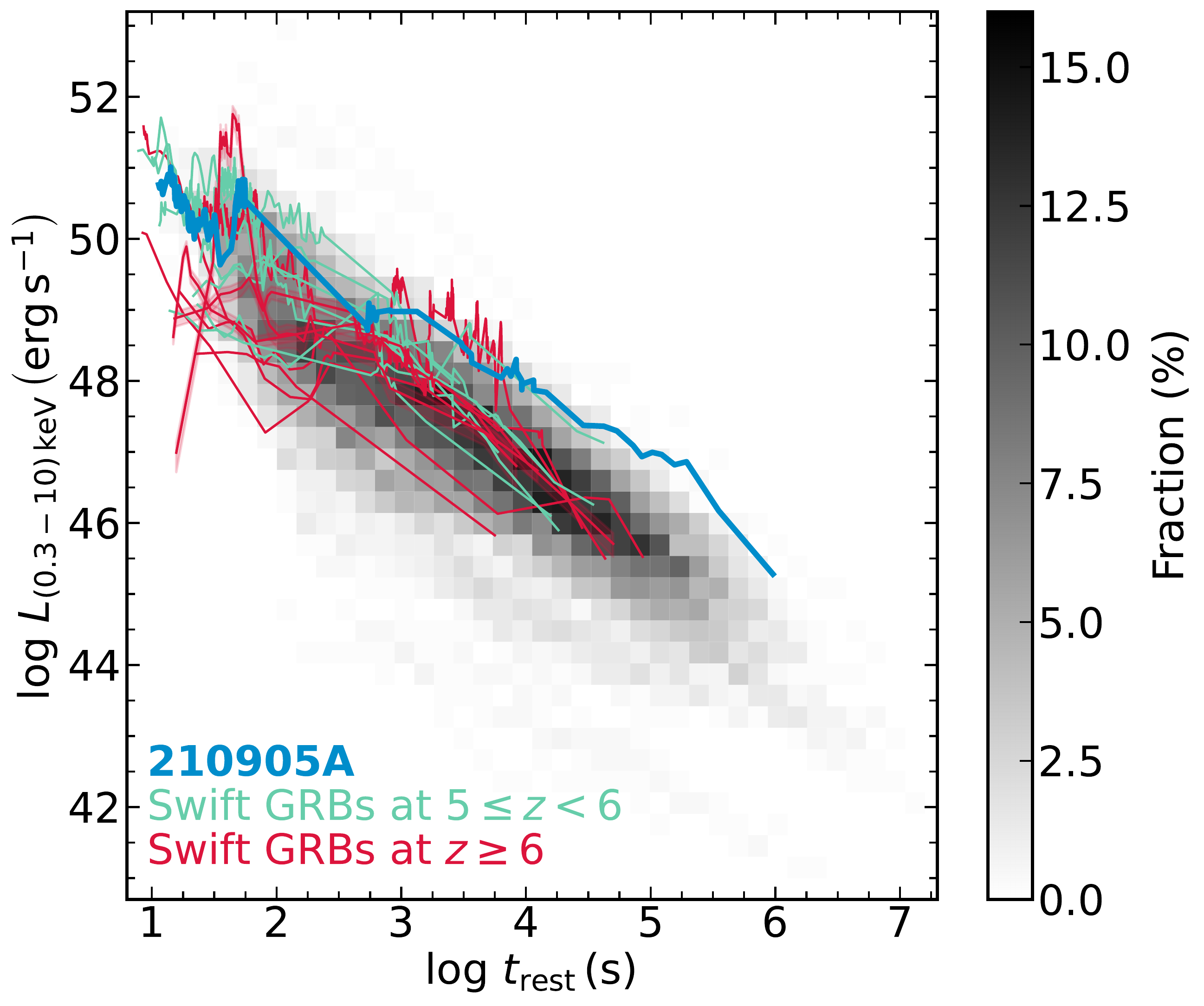}
\caption{
X-ray afterglow of GRB 210905A (blue line) in the context of other high-redshift GRBs (green and red) and the world-sample of \swift{} GRBs with known redshifts (grey density plot). The afterglow of GRB 210905A is the most luminous after 10 ks among all $z>5$ GRBs and one of the most luminous in general. The colour table on the right side translates a grey shade at a given luminosity and time into a fraction of bursts. 
}
\label{fig:schulze}
\end{figure}

To put the X-ray emission in the context of other GRB afterglows, in particular high-redshift GRBs, we retrieved from the \swift{} Burst Analyser
\citep{Evans2010a} the X-ray light curves of 421 long-duration \swift{} GRBs with detected X-ray afterglows (detected in at least two epochs) and known spectroscopic redshifts, which were discovered before the end of July 2022. We processed the data and moved them to their rest-frames following \citet{Schulze2014a}. 
Figure \ref{fig:schulze} shows the parameter space occupied by long-duration GRBs as a density plot and the X-ray light curve of GRB 210905A in blue.
We have also included the X-ray light curves of the high-redshift GRBs 090423, 090429B and 100905A that have only a photometric redshift. The uncertainty in luminosity for these three bursts is indicated by red-shaded regions around the light curves at their redshifts.

 GRB 210905A's X-ray afterglow is among the most luminous at all times. Even compared to other GRBs at $5<z<6$ (green; GRBs 060522, 060927, 130606A, 131227A, 140304A, 201221A, 220521A) and $z>6$ (red; 050904, 080913, 090423, 090429B, 100905A, 120521C, 120923A, 140515A), GRB 210905A has an exceptionally high luminosity. Furthermore, its X-ray afterglow is fading slower than those of most GRBs, at least until the jet break at $\sim5\times10^5$ s in the rest frame (\S\ref{sec:break}).
 Here we note that some of the other bursts at high-$z$ 
 do not show a clear light-curve break in X-rays (GRBs 050904, 080913, 090423, 130606A), although some of them show a break in the optical (GRBs 050904, 090423, 090429B, 120521C), and GRB 140515A  has just one single detection that suggests a possible break similar to GRB 210905A. This is because of the low observed flux of these very high-$z$ afterglows, as only the most luminous events are bright enough for \swift/XRT.
 

\subsection{The optical/NIR afterglow in context}

Following the method devised by \cite{Kann2006ApJ}, we are able to put the NIR afterglow into the context of the (optical/NIR) total afterglow sample. We derive the observer-frame $R_C$ magnitude by shifting all data to the $H$ band, then extrapolating the spectral slope into the observer-frame $R_C$ band, which is completely suppressed at the redshift of the GRB (assuming that there would be no Lyman absorption).
The spectral slope, redshift, and the lack of extinction are then used to derive the magnitude shift $dRc=-5.12^{+0.20}_{-0.21}$ mag to $z=1$. 
The derived $R_C$-band light curve still represents an observed magnitude,  it is as if the GRB were at $z=1$ in a completely transparent universe.

\begin{figure}
\begin{center}
\includegraphics[width=\columnwidth,angle=0]{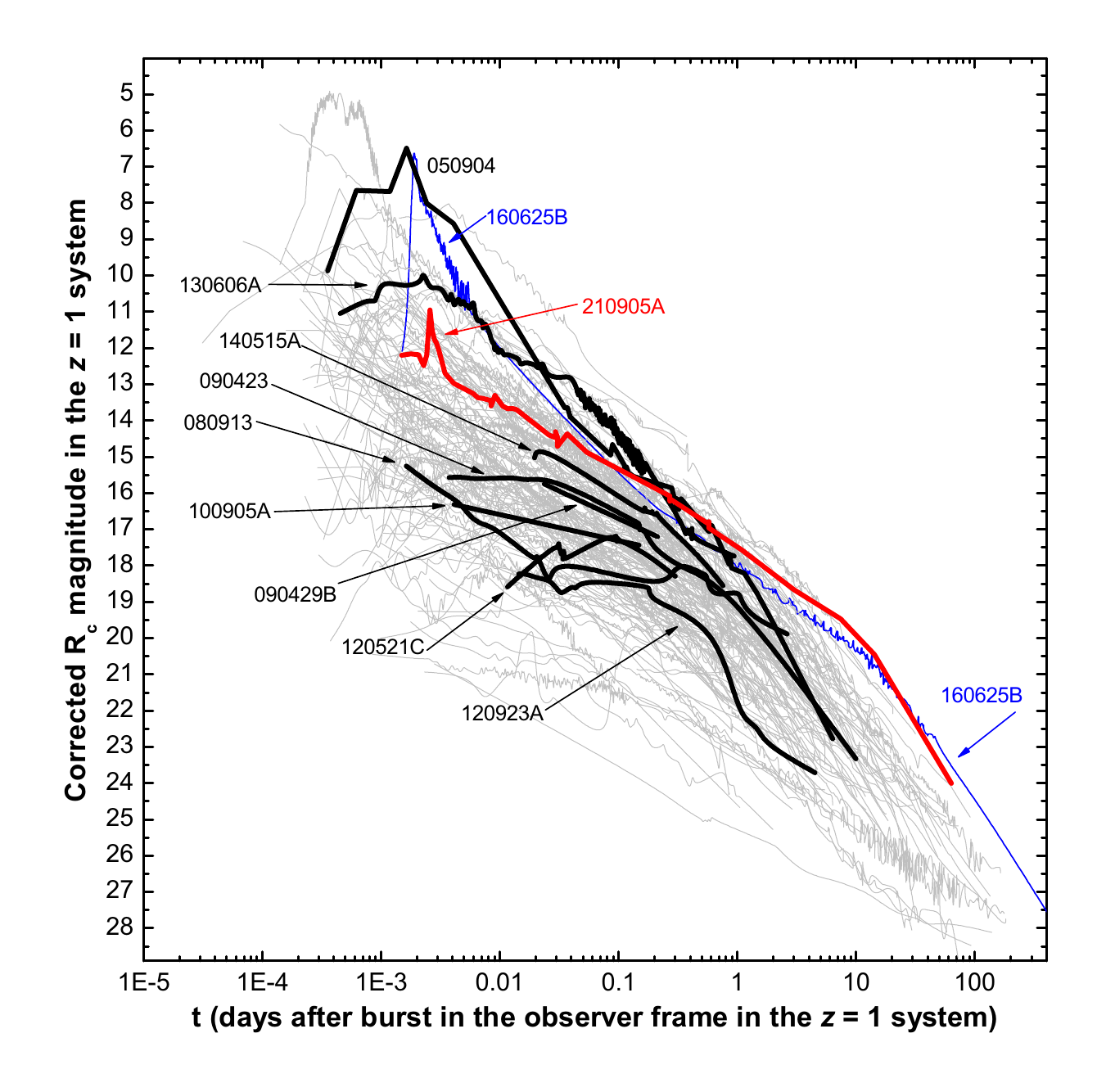}
\caption{The optical afterglow of GRB 210905A (red line) compared to a sample of extinction-corrected afterglows which have all been shifted to $z=1$, from \citet[][2022a,b,c, in prep.]{Kann2006ApJ,Kann2010a,Kann2011a}. Hereby, time and magnitudes are given in the observer frame, but assuming all GRBs are at $z=1$ in a perfectly transparent universe.  
Light grey are LGRBs, thicker black lines GRBs with redshifts $z\gtrsim6$. All magnitudes are in the Vega system. The afterglow of GRB 210905A is the most luminous afterglow ever detected at moderately late times, before finally decaying faster than that of GRB 160625B (blue line). For this light curve, the potential constant source has been subtracted, see \S\ref{sec:constant} for more details. The late-time break in the light curve is clearly visible.} 
\label{fig:kann}%
\end{center}
\end{figure}

We then compare the afterglow with the GRB afterglow light curve samples of \cite{Kann2006ApJ,Kann2010a,Kann2011a} as well as samples from upcoming publications (Kann et al., 2022a,b,c, in prep.). The result is shown in Figure \ref{fig:kann}, with GRB 210905A highlighted in red. The sample of Kann et al. (2022a), in prep. focuses on $z\gtrsim6$ GRBs, and these light curves are highlighted as thick black curves. At early times, the afterglow of GRB 210905A is seen to be among the most luminous known, albeit still fainter than the early afterglows of high-$z$ GRBs 130606A and especially 050904 \citep{Kann2007AJ}. Interestingly, the early flash of GRB 210905A aligns well in rest-frame time (between 70 and 110 s) with those seen in GRB 050904 \citep{Boer2006a}, GRB 160625B \cite[][an extremely energetic lower-redshift GRB, highlighted in blue]{Troja2017Nature}, and, with less contrast, in GRB 130606A \citep{Castro-Tirado2013a}. On the other hand, several bright prompt-associated flashes happen significantly earlier, such as the cases of GRB 080319B \citep{Racusin2008Nature,Bloom2009ApJ} and GRB 120711A \citep[][Kann et al. 2022a, in prep.]{Martin-Carrillo2014a}. Therefore, this similarity in time is likely just a chance coincidence.

An interesting result is found towards the end of the light curve. After removing the potential constant component, the combination of a late break and an early shallow decay makes the afterglow of this burst the most luminous ever detected for a certain time span, before the shallower post-break decay of the afterglow of GRB 160625B (which itself had a very late jet break, \citealt{Kangas2020ApJ}) makes the latter the most luminous known at very late times again (Kann et al. 2022c, in prep.). This provides further evidence for the extremely energetic nature of GRB 210905A.

\subsection{Dust absorption and equivalent hydrogen column densities}\label{sec:avnh}

As in other high-$z$ bursts \citep[see e.g.][]{Zafar2010a,Zafar2011a,Zafar2018a, Melandri2015a}, GRB 210905A is characterised by negligible absorption in the optical/NIR, in agreement with those expected for high-$z$ galaxies populating the faint end of the luminosity function  \citep[e.g.][]{Salvaterra2011a}. In particular,
\cite{McGuire2016a} studied three $z>5.9$ GRB hosts and noted that afterglow analyses in each case pointed to low line-of-sight dust extinction.

Although a low $A_V$ is expected to correlate with a low $N_{H.X}$, the high $N_{H.X}$ value of $7.7^{+3.6}_{-3.2}\times10^{22}\,  \textnormal{cm}^{-2}$ is also not exceptional. It is also observed in other environments, for example in AGNs, and can be naturally explained by the absorption of intervening metals along the line-of-sight \citep[][]{Starling2013a,Campana2015a}, which reside almost entirely in the neutral gas at $z>4.5$ \citep[e.g. ][]{PerouxHowk2020a}, although one cannot exclude the contribution of increasing gas density in the vicinity of the GRB \citep{Heintz2018e}.
The high  $N_{H.X}$ is also in contrast with the  $N_{\rm H\,{\sc I}} \simeq 1.35\times10^{21}\; \textrm{cm}^{-2}$
measured via the Lyman-$\alpha$ absorption-line 
by \cite{Fausey2022a}. The difference can be explained by the very high number of ionising photons produced by the GRB that could ionise the IGM along the line-of-sight up to several hundreds of pc \citep{Saccardi2022a}. 
We discuss the IGM contribution in more detail in \cite{Fausey2022a}.

\subsection{X-ray afterglow luminosity versus prompt energy}\label{sec:lxeiso}

The X-ray luminosity and the isotropic gamma-ray energy release seem to broadly follow a linear relation as already shown by \cite{dePasquale2006a} \citep[see also ][]{Nysewander2009ApJ}, suggesting a roughly universal efficiency for converting a fraction of the initial kinetic energy\footnote{Not to be confused with $E_\mathrm{kin}$, which is the energy left after the prompt phase.} into gamma-ray photons. This was later further confirmed by \cite{Davanzo2012a}. GRBs at $z>6$ also follow this relation.
We test here whether GRB 210905A follows this relation despite its luminosity. We estimate the afterglow X-ray integral flux in the $2-10$ keV rest-frame common energy band and compute the corresponding rest-frame X-ray luminosity at different rest-frame times. The $2-10$ keV rest-frame flux was computed from the observed integral $0.3-10$ keV unabsorbed fluxes and the measured photon index, $\Gamma$ (which we retrieved from the online \swift{} Burst Analyser,
\citealt{Evans2009a,Evans2010a}) in the following way \citep[see][]{Gehrels2008ApJ,Davanzo2012a}:  

\begin{equation} 
f_{X,rf}(2-10 \, {\rm{keV}}) = f_X(0.3-10 \, \rm{keV})\frac{\left({\frac{10}{1+z}}\right)^{2-\Gamma}-\left({\frac{2}{1+z}}\right)^{2-\Gamma}}{{10}^{2-\Gamma}-{0.3}^{2-\Gamma}} \,.
\label{kcorr_eq}
\end{equation}

\begin{figure}
\centering
\includegraphics[width=\columnwidth,angle=0]{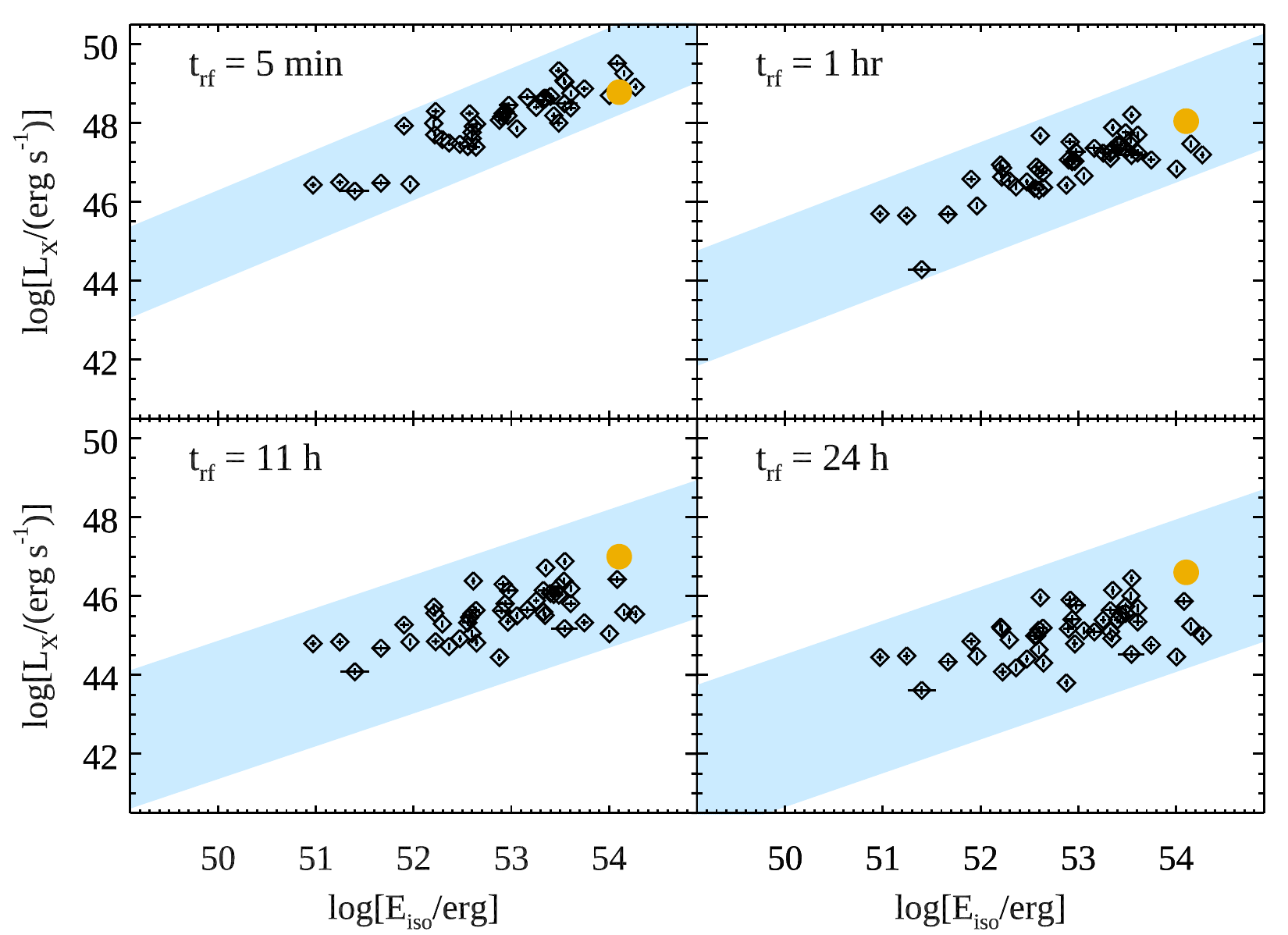}
\caption{The $L_X-E_{\mathrm{iso}}$ correlation presented in \cite{Davanzo2012a} for the long GRBs of the BAT6 sample (diamonds) at different rest-frame times. The shaded area represents the $3\sigma$ scatter of the correlations. GRB 210905A is marked with a filled orange circle.}
\label{fig:lxeiso}
\end{figure}

\noindent The obtained X-ray light curve was then fitted with a multiply broken power-law, after removing the time intervals showing significant flaring, and then the fits were interpolated or extrapolated to the rest-frame times $t_{rf}= 5$ min, $t_{rf}= 1$ hr, $t_{rf}= 11$ hr, $t_{rf}= 24$ hr. 
As shown in Figure~\ref{fig:lxeiso} the properties of GRB\,210905A are fully consistent with the $E_\textrm{iso} - L_{X}$ correlations found for long GRBs by \cite{Davanzo2012a}.

\subsection{Long-GRB progenitors at high redshift}

At high redshift the Universe is expected to be populated by 
pop-III stars, the first stars that formed out of gas clouds of pristine composition.
 Chemical feedback from the supernova explosions of these very massive stars produces metal enrichment within star-forming clouds, raising the metallicity above a critical threshold above which we expect a slow transition of the SFR from massive pop-III to solar-size pop-II and pop-I stars \citep[e.g.][]{Schneider2006a,Maio2010a}.
 Determining how this transition takes place is one of the main missing ingredients to understand galaxy formation in the early Universe. 
 All models  \citep[e.g.][]{MeszarosRees2010a,Toma2011a,Piro2014a} predict pop-III GRBs to be very energetic events, and with very long intrinsic durations of $10^4$ s, making their detection possible even at the highest redshifts. In particular, \cite{Toma2011a} suggested that they can release an equivalent  isotropic energy up to $\sim10^{56-57}$ erg. 

\begin{figure*}
\centering
\includegraphics[width=0.4\textwidth,angle=0]{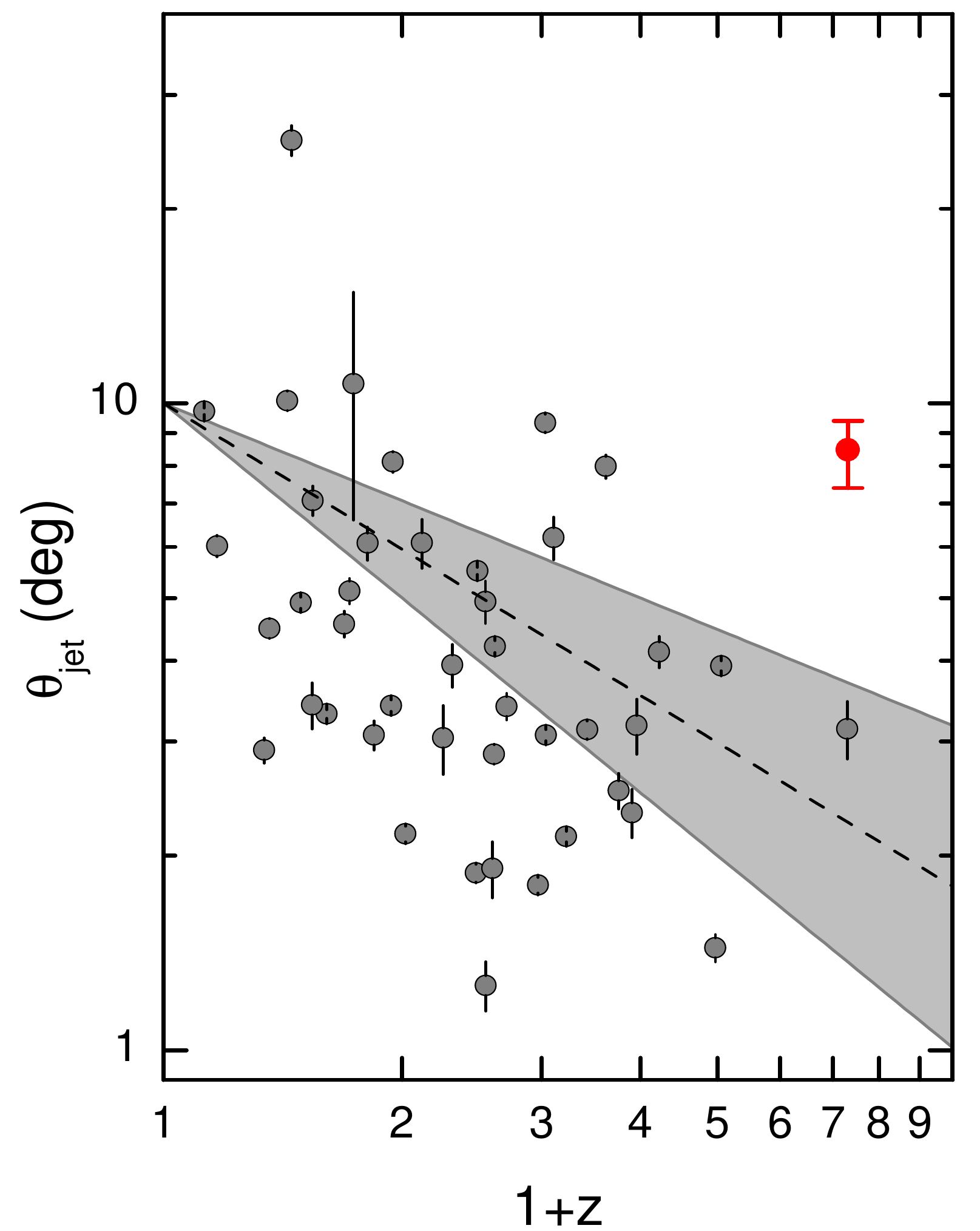}
\includegraphics[width=0.41\textwidth,angle=0]{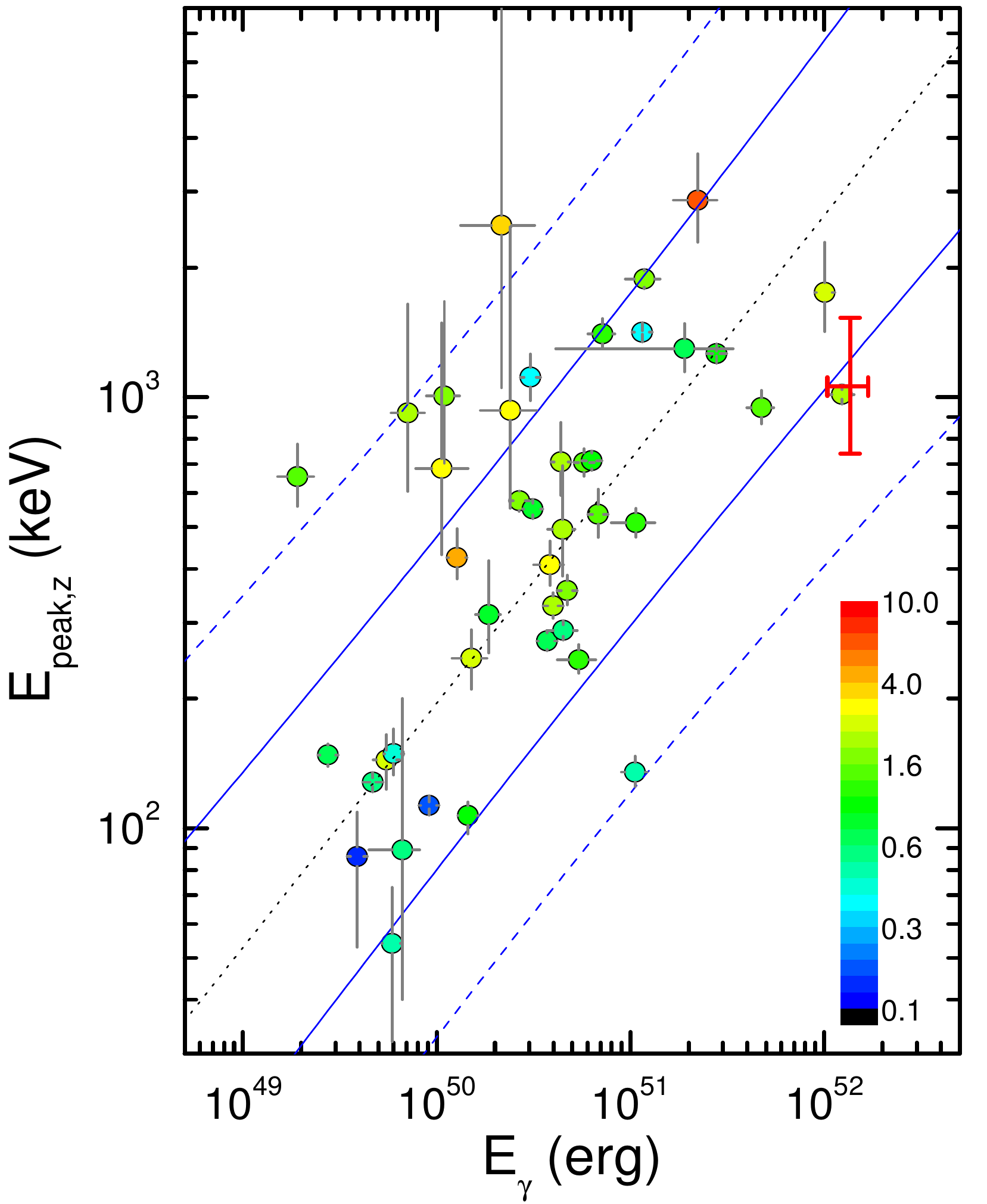}
\caption{Collimated parameters of GRB 210905A (red symbols) compared to a KW sample of 43 long GRBs from \cite{Tsvetkova2017,Tsvetkova2021}. 
We assumed $\eta=0.2$ and $n=1\,\rm cm^{-3}$ for all bursts.
\textit{Left}: Half-opening angle $\theta_\textrm{jet}$  versus redshift. The dashed line within the grey area shows the relation found in \cite{Lloyd-Ronning2020a} with its error. \textit{Right}: $E_\gamma-E_{\mathrm{peak,}z}$ diagram.
As in Figure~\ref{fig:amati}, the colour of each data point represents the burst's redshift. The `Ghirlanda' relation is plotted together with its  68\% and 90\% PIs.
}
\label{fig:collim}
\end{figure*}

In Figure \ref{fig:collim} we compare $\theta_{\rm jet}$ and the collimated energy $E_{\rm \gamma}$ of GRB 210905A with the KW sample of 43 long GRBs with reliable jet-break time estimates \citep{Tsvetkova2017,Tsvetkova2021}.
Considering the uncertainty on the collimation-corrected energy, GRB 210905A 
lies just outside the $1\sigma$ confidence level of the 
$E_{\mathrm{peak,}z}-E_{\rm \gamma}$
 \citep[`Ghirlanda' relation, see ][]{Ghirlanda2004a,Ghirlanda2007a} and thus well compatible with this relation. 
 The energy values involved in GRB 210905A, both isotropic and collimated, are large but do not significantly differ from those at low redshift (see Figs.~\ref{fig:amati} and \ref{fig:collim}). At lower $z$, other events have produced $E\gtrsim10^{54}$~erg isotropically and $E_{\gamma} \simeq 10^{52}$~erg collimation-corrected \citep[see also][]{Cenko2011a}. The most outstanding example is GRB 130427A at $z=0.3399$ (see \S\ref{sec:ori-prompt}), the most powerful GRB at $z<0.9$ \citep[e.g.][]{Maselli2014a,dePasquale2016a}.

GRB 210905A has the highest $E_\gamma$ in the \kw{} catalogue. This and the large $E_{\mathrm{p,}z}$ 
suggest a large bulk Lorentz factor $\Gamma_0$ of the jet. The afterglow light curve, as reported in Figure \ref{fig:lcoptx},
decays as a power-law in both the optical and X-ray band from $\gtrsim5000$~s onwards (observer frame). This suggests that the afterglow deceleration time happened before this epoch. Following the method in \cite{Molinari2007AA}, an upper limit on this peak time provides a lower limit to the maximum bulk Lorentz factor of the jet\footnote{To derive the peak time, we assumed smoothness $k=1$ and decay indices $\alpha_0=-0.7$, $\alpha_0=\alpha_\mathrm{1,opt}=0.69$, before and after the peak, respectively.}, namely $\Gamma_0 \gtrsim 200$, assuming a constant density medium $n_0=1 \;\mathrm{cm}^{-3}$ and the isotropic energy of GRB 210905A (\S\ref{sec:con-prompt}). With this estimate, and the inferred half-opening angle, the burst is consistent with the $\theta_\textrm{jet} - \Gamma_0$ broad anti-correlation reported in \citet[][see their Figure 4]{Ghirlanda2012a}.

Therefore, GRB 210905A, although extremely bright, is not separated markedly from other classical GRBs at low redshift. 
In summary, no features of this event point to a pop-III origin.


\subsection{Star-formation rate at very high redshift}

The rate of GRBs can be used to estimate the SFR in the remote Universe (see \S\ref{sec:intro}).
Recently, \citet[and references therein]{Lloyd-Ronning2019a,Lloyd-Ronning2020a,Lloyd-Ronning2020b} have argued
that at high redshift, the 
GRB jets were,
on average, narrower than those of closer GRBs \citep[see also][]{Laskar2014a,Laskar2018a}. This would imply that more stars formed at high redshift than previously estimated, unless the GRB properties, and thus their rate, are extremely environment-sensitive \citep{Kistler2008a,Kistler2009a,Robertson2012a,Jakobsson2012a,Tanvir2012a,Japeli2016a,Palmerio2019a}.
In the left panel of Figure~\ref{fig:collim}, we report the relation found by the above authors in the $\theta_\mathrm{jet}-(1+z)$ plane. GRB 210905A is an outlier event located at $2-3\sigma$ above this relation. We observe that the half-opening angle of this GRB at $z = 6.312$  ($\theta_{\rm jet}\sim 8$ deg) is 
consistent with the median value of $\theta_{\rm jet}=7.4_{-6.6}^{+11}$ deg for GRBs at $z\sim1$ but larger than the mean of $\theta\sim3.6\pm0.7$ deg found using three $z>6$ bursts \citep[GRBs 050904, 090423, 120521C,][]{Laskar2014a,Laskar2018a}. When we include GRB 210905A, the mean for the $z\gtrsim6$ bursts is $\theta\sim4.8\pm0.6$ deg, closer to the best value for $z\sim1$ events.
These findings would argue against a putative inverse correlation between $z$ and $\theta_{\rm jet}$.


\section{Conclusions}\label{sec:con}

GRB 210905A was a long burst at redshift $z=6.312$. 
Our extensive and prompt follow-up observations from optical/NIR to X-ray and gamma-ray bands, starting in the first seconds, have allowed us to study in detail both the prompt and the afterglow phases. 
 We carried out a joint time-resolved analysis of the last of the three pulses of the prompt emission, which is shown to be in agreement with synchrotron emission, similar to other bursts at lower redshifts. 
Among the sample of ten $z\gtrsim6$ GRBs known to date, GRB 210905A
stands out (together with GRB 050904), having the highest isotropic energy release and among the highest afterglow luminosity at late times, while still being consistent with the range of values found for other long GRBs. 

The temporal evolution of the afterglow can be interpreted as due to energy injection followed by a decay well in agreement with the slow-cooling scenario and a constant-density (`ISM') circumburst medium profile within the standard fireball theory.
However, the optical and X-ray afterglows are among the most luminous ever detected, in particular in the optical range at $t\gtrsim0.5$ d in the rest frame, due to very slow fading and a late jet break. 
In late \textit{HST} imaging, we find evidence for an 
underlying host with UV luminosity slightly larger
than that of galaxies contributing the most to star formation at $z=6-7$.
If confirmed with further observations,
the host of GRB 210905A would be the fourth and the brightest GRB host at $z>6$ detected to date. It would also be bright enough to be characterised via spectroscopy with the \textit{JWST} \citep[e.g.][]{McGuire2016a}, providing one of the first and better estimates on the SFR, metallicity and dust content of a GRB host at very high redshift.

The jet break at $\sim50$ d (observer frame) results in a half-opening angle that is larger than that of other $z>6$ bursts, thus putting into question the putative inverse dependence of the half-opening angle on redshift. 
The large total energy budget of $E_{\rm total}>10^{52}$ erg  associated with this GRB likely excludes all but the most extreme magnetar models as a central engine of this GRB. Therefore, our analysis  leaves the Kerr black hole as the preferred scenario for the central engine of GRB 210905A.
Finally, the shallow evolution before 1 day suggests that the black hole injected energy via stratified mass ejecta with different Lorentz factors.

In summary, this burst is consistent with the `Amati', `Ghirlanda', and  `Yonetoku' relations. This fact, and the agreement with the $E_{\rm iso} - L_{X}$ plane show that GRB 210905A is a very energetic event but still in the upper tail of the prompt energy and X-ray luminosity distributions of long GRBs. It is not unexpected that our view of the high-$z$ GRB Universe is biased towards the most luminous events, simply because our instruments are limited in sensitivity. In other words, despite its outstanding luminosity it is unlikely that the origin of this GRB is different from those of low-redshift GRBs, such as a pop-III progenitor.

Gamma-ray bursts at $z\gtrsim6$ are rare events from the perspective of today's follow-up capabilities, but they are just a small part of a larger population that future proposed missions promise to uncover (e.g. \textit{THESEUS}, \citealt{Amati2018a}; \textit{Gamow}, \citealt{White2021a}) and, in synergy with the largest ground- and space-based telescopes (such as the James Webb Space Telescope), to answer open questions in modern astrophysics such as 
the identification of the sources responsible for cosmic reionisation, and the evolution of SFR and metallicity across the transition from pop-III stars to pop-II and pop-I stars.

\begin{acknowledgements}

We thank the anonymous referee for providing  thoughtful comments.

We acknowledge useful discussion with L. Nicastro and A. MacFadyen.

A. Rossi acknowledges support from the INAF project Premiale Supporto Arizona \& Italia.

D.D.F. and A.E.T. acknowledge support from RSF grant 21-12-00250.

D.A.K. acknowledges support from Spanish National Research Project RTI2018-098104-J-I00 (GRBPhot).

A.R., E.Pal., P.D.A., L.A., E.Pi., G.S., S.C., V.D.E., M.D.V., and A.M. acknowledge support from PRIN-MIUR 2017 (grant 20179ZF5KS).

P.D.A., A.M. acknowledge support from the Italian Space Agency, contract ASI/INAF n. I/004/11/5.

L.I. was supported by grants from VILLUM FONDEN (project number 16599 and 25501).

D.B.M. and A.J.L. acknowledge the European Research Council (ERC) under the European Union's Seventh Framework programme (FP7-2007-2013) (Grant agreement No. 725246). The Cosmic Dawn Center (DAWN) is funded by the Danish National Research Foundation under grant No. 140. 

K.E.H. acknowledges support by a Postdoctoral Fellowship Grant (217690--051) from The Icelandic Research Fund. 

C.G.M. acknowledges financial support from Hiroko and Jim Sherwin.
 
Part of the funding for GROND (both hardware as well as personnel) was generously granted from the Leibniz-Prize to Prof. G. Hasinger (DFG grant HA 1850/28-1).
 
This work made use of data supplied by the UK \swift{} Science Data Centre at the University of Leicester.

\end{acknowledgements}


\bibliographystyle{aa}
\bibliography{biblio} 

\end{CJK*}
\end{document}

%% file: authaa.tex
   \author{A. Rossi           \inst{1}
          \fnmsep\thanks{E-mail:andrea.rossi@inaf.it}
          \and
D.~D.~Frederiks\inst{2}
          \and
D.~A. Kann \inst{3} 
        \and
M. De~Pasquale \inst{4}
        \and
E. Pian \inst{1}
    \and        
 G. Lamb \inst{5}
    \and    
P. D'Avanzo  \inst{6}
\and
L. Izzo  \inst{7}
\and
A.~J. Levan \inst{8}
    \and
 D.~B. Malesani \inst{8,9,10}
    \and
 A. Melandri  \inst{6}
\and
A. Nicuesa Guelbenzu  \inst{11}
\and
S. Schulze \inst{12}
\and
R. Strausbaugh \inst{13}
\and
N.~R. Tanvir  \inst{5}
\and
L. Amati \inst{1}
\and
 S. Campana  \inst{6}
\and
A. Cucchiara \inst{13,14}
\and
 G. Ghirlanda  \inst{5,15}
 \and
 M. Della Valle \inst{16}
\and
 S. Klose   \inst{11}
\and
R. Salvaterra \inst{17}
\and
R.~L.~C. Starling \inst{5}
\and
G. Stratta \inst{1,18}
\and
A.~E.~Tsvetkova \inst{2}
\and
S.~D. Vergani \inst{5,19,20}
\and
A. D'A\`i \inst{21}
\and
D. Burgarella \inst{22}
\and
S. Covino \inst{6}
\and
V. D'Elia \inst{23,24}
\and
A. de Ugarte Postigo \inst{25}
\and
H. Fausey \inst{26}
\and
J.~P.~U.~Fynbo \inst{9,10}
\and
F. Frontera \inst{1,27}
\and
C. Guidorzi \inst{1,27,28}
\and
K.~E.~Heintz \inst{9,10,29}
\and
N. Masetti \inst{1,30}
\and
E. Maiorano \inst{1}
\and
C.~G. Mundell \inst{31}
\and
S.~R. Oates \inst{32}
\and
M.~J. Page \inst{33}
\and
E. Palazzi \inst{1}
\and
J. Palmerio \inst{19}
\and
G. Pugliese \inst{34,35}
\and
A. Rau \inst{36}
\and
A. Saccardi \inst{19}
\and
B. Sbarufatti \inst{5,37}
\and
D.~S. Svinkin\inst{2}
\and
G. Tagliaferri \inst{6}
\and
A.~J. van der Horst \inst{26,38}
    \and
D.~J. Watson \inst{6,9,10}
\and
M.~V. Ulanov\inst{2}
\and
K. Wiersema \inst{39}
\and
D. Xu \inst{40,41}
\and
J.~Zhang (张洁莱)\inst{42,43}
          }
   \institute{INAF - Osservatorio di Astrofisica e Scienza dello Spazio, Via Piero Gobetti 93/3, 40129 Bologna,\, Italy 
   \and
   Ioffe Institute, Politekhnicheskaya 26, St.~Petersburg 194021, Russia 
   \and
   Instituto de Astrof\'isica de Andaluc\'ia (IAA-CSIC), Glorieta de la Astronom\'ia s/n, 18008 Granada, Spain 
   \and
   University of Messina, MIFT Department, Polo Papardo, Viale F.S. D'Alcontres 31, 98166 Messina, Italy 
      \and
   School of Physics and Astronomy, University of Leicester, University Road, Leicester LE1 7RH, UK 
   \and
   INAF - Osservatorio Astronomico di Brera, Via E. Bianchi 46, I-23807, Merate (LC), Italy 
   \and
   DARK, Niels Bohr Institute, University of Copenhagen, Jagtvej 128, 2200 Copenhagen, Denmark 
  \and 
    Department of Astrophysics/IMAPP, Radboud University, 6525 AJ Nijmegen, The Netherlands 
    \and
    Cosmic DAWN Center, Denmark 
\and
    Niels Bohr Institute, University of Copenhagen, Jagtvej 128, DK-2200 Copenhagen, Denmark 
   \and
   Th\"uringer Landessternwarte Tautenburg, Sternwarte 5, 07778 Tautenburg, Germany 
   \and
   The Oskar Klein Centre, Physics Department of Physics, Stockholm University, Albanova University Center, SE 106 91 Stockholm, Sweden  
   \and
   University of the Virgin Islands, Number 2 Brewers Bay Rd., St. Thomas, VI 00802, USA 
   \and
   College of Marin, 120 Kent Avenue, Kentfield CA 94904 
   \and
   INFN - Sezione di Milano-Bicocca, Piazza della Scienza 3, I-20126 Milano, Italy 
   \and 
    INAF, Osservatorio Astronomico di Capodimonte, Salita Moiariello 16, 80131 Naples, Italy  
   \and
   INAF - IASF Milano, Via Alfonso Corti 12, I-20133 Milano, Italy 
   \and
   INAF - Istituto di Astrofisica e Planetologia Spaziali, via Fosso del Cavaliere 100, I-00133 Roma, Italy
   \and
   GEPI, Observatoire de Paris, PSL University, CNRS, 5 Place Jules Janssen, 92190 Meudon, France 
   \and
   Institut d’Astrophysique de Paris, UMR 7095, CNRS-SU, 98 bis
   boulevard Arago, 75014, Paris, France
  \and
   INAF - IASF Palermo, via Ugo La Malfa 153, I-90146 - Palermo, Italy 
   \and 
Aix Marseille Univ, CNRS, CNES, LAM, Marseille, France 
   \and
Space Science Data Center (SSDC) - Agenzia Spaziale Italiana (ASI), I-00133 Roma, Italy 
\and
INAF - Osservatorio Astronomico di Roma, Via Frascati 33, 00040 Monte Porzio Catone, Italy
\and
Artemis, Observatoire de la C\^ote d'Azur, Universit\'e C\^ote d'Azur, CNRS, 06304 Nice, France
\and
Department of Physics, The George Washington University, 725 21st Street NW, Washington, DC 20052, USA  
\and
Department of Physics and Earth Science, University of Ferrara, via Saragat 1, I-44122, Ferrara, Italy  
\and 
INFN – Sezione di Ferrara, Via Saragat 1, 44122 Ferrara, Italy  
\and
Centre for Astrophysics and Cosmology, Science Institute, University of Iceland, Dunhagi 5, 107, Reykjavík, Iceland 
    \and
Departamento de Ciencias F\'isicas, Universidad Andr\'es Bello, Fern\'andez Concha 700, Las Condes, Santiago, Chile 
\and 
Department of Physics, University of Bath, Claverton Down, Bath, BA2 7AY 
    \and
    School of Physics and Astronomy and Institute for Gravitational Wave Astronomy, University of Birmingham, Birmingham B15 2TT, UK 
    \and
    Mullard Space Science Laboratory, University College London, Holmbury St Mary, Dorking, Surrey, RH5 6NT, UK 
   \and
Anton Pannekoek Institute for Astronomy, University of Amsterdam, Science Park 904, 1098 XH Amsterdam, The Netherlands 
   \and
Leiden Observatory, University of Leiden, Niels Bohrweg 2, 2333 CA Leiden, The Netherlands   
   \and
   Max-Planck-Institut f\"ur extraterrestrische Physik,
Giessenbachstra{\ss}e 1, 85748 Garching, Germany 
   \and
   Department of Astronomy and Astrophysics, The Pennsylvania State University, University Park, PA 16802, USA  
\newpage
   \and
   Astronomy, Physics, and Statistics Institute of Sciences (APSIS), 725 21st Street NW, Washington, DC 20052, USA 
   \and
   Physics Department, Lancaster University, Lancaster, LA1 4YB, UK
   \and
   Key Laboratory of Space Astronomy and Technology, National Astronomical Observatories, Chinese Academy of Sciences, Beijing 100101, China
   \and 
   School of Astronomy and Space Sciences, University of Chinese Academy of Sciences, 19A Yuquan Road, Beijing 100049, China
   \and
    Centre for Astrophysics and Supercomputing, Swinburne University of Technology, Mail Number H29, PO Box 218, 31122 Hawthorn, VIC, Australia 
\and
ARC Centre of Excellence for Gravitational Wave Discovery (OzGrav), Hawthorn, 3122, Australia 
   }

%% file: texdata.tex
\begin{table}
\centering
\caption{Optical/NIR photometry of the afterglow (AB magnitudes).}
\begin{threeparttable}
\setlength{\tabcolsep}{0.6em}
\small
\begin{tabular}{lcccc} 
\toprule
$\Delta t$\tnote{a} & Magnitude    & Filter   & Telescope/   \\ 
(days)    & AB\tnote{b}           &          &   Instrument           \\  
\midrule  
0.11410 &	$>25.0$ & 	 $g^\prime$ &	VLT/XS	\\ 
0.10930 &	$>24.8$ & 	 $r^\prime$ &	VLT/XS	\\ 
0.03771 &	$>22.5$ 	& 	 $R_C$ &	LCO/SINISTRO	\\ 
0.05867 &	$19.46\pm0.14$ & 	 $I_C$ &	LCO/SINISTRO	\\ 
0.10739 &	$20.35\pm0.14$ & 	 $I_C$ &	LCO/SINISTRO	\\ 
0.13519 &	$20.41\pm0.15$ & 	 $I_C$ &	LCO/SINISTRO	\\ 
0.10627 &	$20.52\pm0.06$ & 	 $I_{\rm Bessel}$ &	VLT/XS	\\ 
87.0326   & $>26.0$& $I_{\rm Bessel}$ &   VLT/FORS2 \\
0.02651 &	$20.86\pm0.46$ & 	 $i^\prime$ &	REM/ROSS	\\ 
0.19712 &	$22.78\pm0.03$ & 	 $i^\prime$ &	Blanco/DECam	\\ 
0.02614 &	$18.17\pm0.33$ & 	 $Z$ &	REM/REMIR	\\ 
0.03065 &	$18.23\pm0.37$ & 	 $z^\prime$ &	REM/ROSS	\\ 
0.05048 &	$18.48\pm0.29$ & 	 $Z$ &	REM/REMIR	\\ 
0.07097 &	$>18.1$ & 	 $z^\prime$ & REM/ROSS \\	
0.10722 &	$19.20\pm0.40$ & 	 $Z$ &	REM/REMIR	\\ 
0.11129 &	$19.53\pm0.02$ & 	 $z^\prime$ &	VLT/XS	\\ 
0.98756 &	$21.61\pm0.12$ & 	 $z^\prime$ &	MPG/GROND	\\ 
2.09740 &	$22.34\pm0.18$ & 	 $z^\prime$ &	MPG/GROND	\\
0.01489 &	$17.12\pm0.15$ & 	 $J$ &	REM/REMIR	\\ 
0.02402 &	$17.52\pm0.16$ & 	 $J$ &	REM/REMIR	\\ 
0.03315 &	$17.45\pm0.19$ & 	 $J$ &	REM/REMIR	\\ 
0.04662 &	$17.81\pm0.15$ & 	 $J$ &	REM/REMIR	\\ 
0.11155 &	$18.86\pm0.22$ & 	 $J$ &	REM/REMIR	\\ 
0.98756 &	$20.26\pm0.11$ & 	 $J$ &	MPG/GROND	\\
2.09740 &	$21.14\pm0.12$ & 	 $J$ &	MPG/GROND	\\
231.815 &	$25.66\pm0.05$ & 	 $F140W$ &	HST/WFC3 NIR	\\
0.00542 &	$16.15\pm0.09$ & 	 $H$ &	REM/REMIR	\\ 
0.00649 &	$16.11\pm0.09$ & 	 $H$ &	REM/REMIR	\\ 
0.00756 &	$16.13\pm0.11$ & 	 $H$ &	REM/REMIR	\\ 
0.00832 &	$16.43\pm0.26$ & 	 $H$ &	REM/REMIR	\\ 
0.00883 &	$16.15\pm0.25$ & 	 $H$ &	REM/REMIR	\\ 
0.00940 &	$14.90\pm0.09$ & 	 $H$ &	REM/REMIR	\\ 
0.00983 &	$15.47\pm0.11$ & 	 $H$ &	REM/REMIR	\\ 
0.01027 &	$15.71\pm0.12$ & 	 $H$ &	REM/REMIR	\\ 
0.01070 &	$15.82\pm0.16$ & 	 $H$ &	REM/REMIR	\\ 
0.01114 &	$16.00\pm0.13$ & 	 $H$ &	REM/REMIR	\\ 
0.02397 &	$17.32\pm0.12$ & 	 $H$ &	REM/REMIR	\\ 
0.03853 &	$17.55\pm0.16$ & 	 $H$ &	REM/REMIR	\\ 
0.09519 &	$18.35\pm0.21$ & 	 $H$ &	REM/REMIR	\\ 
0.98756 &	$20.08\pm0.12$ & 	 $H$ &	MPG/GROND	\\
0.98777 &	$20.08\pm0.06$ & 	 $H$ &	VLT/HAWK-I	\\
2.07128 &	$20.78\pm0.06$ & 	 $H$ &	VLT/HAWK-I	\\
2.0974 &	$20.64\pm0.14$ & 	 $H$ &	MPG/GROND	\\
3.97651 &	$21.49\pm0.06$ & 	 $H$ &	VLT/HAWK-I	\\ 
10.9949 &	$22.56\pm0.07$ & $H$ &	VLT/HAWK-I\\ 
27.0106 &	$23.30\pm0.12$ & $H$ &	VLT/HAWK-I\\ 
52.0603 &   $24.14\pm0.13$  & $H$ &   VLT/HAWK-I \\
0.01259 &	$16.16\pm0.15$ & 	 $K$ &	REM/REMIR	\\ 
0.02172 &	$16.72\pm0.24$ & 	 $K$ &	REM/REMIR	\\ 
0.03085 &	$17.07\pm0.28$ & 	 $K$ &	REM/REMIR	\\ 
0.04257 &	$17.14\pm0.23$ & 	 $K$ &	REM/REMIR	\\ 
0.10746 &	$18.33\pm0.38$ & 	 $K$ &	REM/REMIR	\\ 
0.98756 &	$20.17\pm0.20$ & 	 $K_s$ &	MPG/GROND	\\
1.00021 &	$19.89\pm0.03$ & 	 $K_s$ &	VLT/HAWK-I	\\
\bottomrule                                 
\end{tabular}   
\begin{tablenotes}\footnotesize 
\item[a] Mid-time after the burst trigger.
\item[b] The photometry is not corrected for Galactic extinction. 
\end{tablenotes}    
\label{tab:photall}
\end{threeparttable}         
\end{table}